\documentclass[prd,preprintnumbers,superscriptaddress,nofootinbib,twocolumn]{revtex4-2}
\usepackage{epsfig}
\usepackage{amsmath}
\usepackage{amsfonts}
\usepackage{float}
\usepackage{amssymb}
\usepackage{slashed}
\usepackage{color}
\usepackage{pbox}
\usepackage{subfigure}
\usepackage[colorlinks,citecolor=blue, linkcolor = blue, urlcolor = blue]{hyperref}
\usepackage{tabularx}
\usepackage{titlesec}
\usepackage{scrextend}
\usepackage[normalem]{ulem}
\usepackage{orcidlink}

\def\lsim{\raise0.3ex\hbox{$\;<$\kern-0.75em\raise-1.1ex\hbox{$\sim\;$}}}
\def\gsim{\raise0.3ex\hbox{$\;>$\kern-0.75em\raise-1.1ex\hbox{$\sim\;$}}}
\usepackage{MnSymbol}

\def\lsim{\lesssim}
\def\gsim{\gtrsim}

\def\bl{\rm B-L}

\def\vbl{v_{\rm BL}}
\def\zbl{Z_{\rm BL}}
\def\ni{N_i}
\def\gbl{g_{\rm BL}}
\def\ubl{U(1)_{\rm B-L}}
\def\qbl{{\rm Q}_{\rm BL}}
\def\rbl{r_{\rm BL}}
\def\rf{r_{\eta}}
\def\lamf{\lambda_\phi}

\def\yf{Y_i}
\def\nn{\nonumber}
\def\Yd{Y_{\rm D}}
\newcommand{\mbar}{\overline{m}}

\newcommand{\vh}{v_h}

\def\bal#1\eal{\begin{align}#1\end{align}}
\graphicspath{{figures/}}
\begin{document}
\title{Probing low scale leptogenesis through gravitational wave}

\author{Anirban Biswas\,\orcidlink{0000-0002-3810-3326}}
\email{anirban.biswas.sinp@gmail.com}
\affiliation{Department of Physics, Gaya College (a constituent
unit of Magadh University, Bodh Gaya), Gaya 823001, India}

\author{Sougata Ganguly\,\orcidlink{0000-0002-8742-0870}}
\email{sganguly0205@ibs.re.kr}
\affiliation{Particle Theory  and Cosmology Group, Center for Theoretical Physics of the Universe,\\
Institute for Basic Science (IBS), Daejeon, 34126, Korea}

\begin{abstract}
The quest for a common origin of neutrino mass and baryogenesis is
one of the long-standing goals in particle physics. A minimal gauge
extension of the Standard Model by $U(1)_{\rm B-L}$ symmetry provides
a unique scenario to explain the tiny mass of neutrinos as well as the
observed baryon asymmetry, both by virtue of three right-handed
neutrinos (RHNs). Additionally, the $U(1)_{\rm B-L}$ breaking scalar
that generates mass of the RHNs can produce a stochastic gravitational
wave background (SGWB) via cosmological first-order phase transition.
In this work, we systematically investigate TeV-scale leptogenesis
considering flavor effects that are crucial in the low-temperature regime. We also explore all possible RHN production
channels which can have significant impact on the abundance of RHNs,  depending on the value of the $U(1)_{\rm B-L}$ gauge coupling. We demonstrate that
the strong dependence of the $U(1)_{\bl}$ gauge sector
on the baryon asymmetry as well as SGWB production can be utilized
to probe a region of the model parameter space. In particular, 
we find that a $U(1)_{\rm B-L}$
gauge boson with mass $\sim 10\,\rm TeV$ and gauge coupling 
$\sim 0.1$ can explain the observed baryon asymmetry and produces a detectable SGWB in future detectors as well. Importantly, this region falls beyond the reach of the current
collider sensitivity.
 
\end{abstract}
\preprint{CTPU-PTC-25-11}
\maketitle
\section{INTRODUCTION}
\label{sec:introduction}
The Standard Model (SM) of particle physics
is an extremely successful theory in order
to explain the observable phenomena below the TeV scale. Despite its tremendous successes, the observation of neutrino oscillation \cite{Super-Kamiokande:1998kpq, SNO:2002tuh} raises a fundamental question about the completeness of the SM since neutrinos are massless in the SM. Neutrino oscillation predicts that the neutrinos have a very tiny mass, typically in the ballpark of $0.1\,\rm eV$. There are many beyond the Standard Model (BSM) frameworks that explain
the neutrino masses and their mixings. However, the SM augmented with at least two heavy right-handed neutrinos (RHNs) is the simplest
BSM scenario in order to explain the tininess of the neutrino mass as well as the neutrino oscillation parameters, and it is known as the ``type-I seesaw" mechanism \cite{Minkowski:1977sc,Yanagida:1979as}. 

Another important shortcoming of the SM is its inability to explain the observed baryon asymmetry of the Universe. The observation of the cosmic microwave background (CMB) by the {\it Planck} satellite and the prediction of big bang nucleosynthesis strongly suggest an excess of the baryon over the antibaryon, and it is quantified as the 
baryon-to-photon ratio $\eta_B$. The current value of $\eta_B$
is $(6.143 \pm 0.190)\times 10^{-10}$ \cite{ParticleDataGroup:2024cfk}. The explanation of $\eta$ requires charge-parity ({\it CP}) violation in the theory. In the SM, {\it CP} violation exists in the quark sector, 
but the amount of {\it CP} violation (${\cal O} (10^{-20}$)) is way more smaller than the value required to explain the CMB prediction of $\eta_B$ \cite{Davidson:2008bu}.

In light of the baryon asymmetric universe, many baryogenesis mechanisms have been proposed \cite{Yoshimura:1978ex, Ellis:1978xg, Nanopoulos:1979gx, Affleck:1984fy, Dine:1995kz,Riotto:1999yt, Dine:2003ax, Cline:2006ts, Morrissey:2012db}. 
Among them, the production of baryon asymmetry from the asymmetry in the leptonic sector known as ``leptogenesis" is a very
well-studied scenario \cite{Luty:1992un, Gherghetta:1993kn, Plumacher:1996kc, Covi:1996wh, Plumacher:1997ru, Plumacher:1998ex, Buchmuller:1999cu, Buchmuller:2000as, Buchmuller:2004nz, Giudice:2003jh, Davidson:2008bu,Iso:2010mv, Sahu:2005fe, Geng:2017foe, Biswas:2017tce, Pramanick:2022put, Pramanick:2024gvu, 
Das:2024gua,
Bose:2024bnp}. In the SM, the baryon number $\rm B$ and the lepton number $\rm L$ are not good quantum numbers since they are not conserved at the quantum level. However, $\bl$ is a good quantum number. The basic principle of leptogenesis is to
convert the asymmetry in the leptonic sector to the asymmetry in the baryonic sector via the $\bl$ conserving
sphaleron process which is active when the temperature of the Universe ($T$) lies between $\sim 130$ and $10^{12}\,\rm GeV$. Thus, asymmetry in the leptonic sector, produced in the temperature range between $130$ and $10^{12}\,\rm GeV$ can be efficiently converted into baryon asymmetry, --thanks to $\bl$ conserving and $\rm B+L$ violating the electroweak sphaleron process. 

The production of baryon asymmetry via leptogenesis is a natural consequence of the type-I seesaw framework.
In this framework, {\it CP} violation in the leptonic sector is sourced by the complex Yukawa couplings
between the heavy RHNs and the SM sectors. In the presence of complex Yukawa couplings, the decay of heavy RHNs
into SM leptons and the Higgs doublet generates the {\it CP} asymmetry. Under the assumption that the heavy RHNs 
with the hierarchical mass spectrum are in thermal equilibrium in the early Universe, 
the lepton asymmetry generated by the lightest RHN only survives. In this vanilla leptogenesis scenario, the mass of the RHN is required to 
be larger than $10^{9}\,\rm GeV$, known as the Davidson-Ibarra (DI) bound \cite{Davidson:2002qv}, in order to explain the baryon asymmetry.

The minimum mass required in this scenario is much larger than the current reach of collider experiments. 
The resonant leptogenesis is an interesting possibility to lower the minimum mass of the RHN to approximately the TeV scale. 
In this mechanism, the key idea is that if the two RHNs have a nearly degenerate mass spectrum, the leptonic asymmetry will 
be resonantly enhanced and generate the correct baryon asymmetry via leptogenesis, even if the mass of the RHNs is in the TeV scale. 
However, another important aspect of low scale leptogenesis is the flavor effects. When the temperature of the Universe is 
higher than $10^{12}$ GeV, the Yukawa interactions of the SM leptons are out of equilibrium, making
different lepton flavors indistinguishable. However, as the temperature becomes smaller than $10^{12}$ GeV, the Yukawa interaction involving 
a tau lepton enters the equilibrium. As a result, for $10^{9} \,{\rm GeV}\lesssim T\lesssim 10^{12}\,\rm GeV$,
the tau flavor can be discriminated from the other two flavors. Moreover,  for $T\lesssim 10^9$ GeV, the Yukawa interaction of 
the second-generation leptons also enters the equilibrium and muon flavor is also distinguishable. 
Finally, when the temperature $T\lesssim10^5$ GeV, all three flavors are in thermal equilibrium through Yukawa interactions, and 
we need to consider their effects in the lepton asymmetry calculation separately, as discussed in 
\cite{Abada:2006ea, Nardi:2006fx, Blanchet:2006be, Adhikary:2014qba, AristizabalSierra:2014nzr, Dev:2017trv}.

Baryon asymmetry produced via leptogenesis requires a new physics at a scale much larger than the electroweak scale. 
The detection of the stochastic gravitational wave background (SGWB) in future gravitational wave (GW) detectors can be used to probe these kinds of high scale theories.
The SGWB in the early Universe can originate from cosmic strings \cite{Vilenkin:1981bx,Vachaspati:1984gt,Damour:2004kw,
Schmitz:2020syl}, quantum fluctuations
during reheating \cite{Grishchuk:1974ny, Grishchuk:1993ds}, first-order phase transitions 
\cite{Witten:1984rs, Kamionkowski:1993fg}, etc. (see \cite{Maggiore:1999vm} for more details). BSM models with an 
extended scalar sector are an obvious source of SGWB production in the early Universe via first-order phase transition 
(FOPT) and it has been extensively studied in various BSM models, as discussed in \cite{Cline:1996mga,
Profumo:2007wc, Barger:2008jx,  FileviezPerez:2008bj, Gonderinger:2009jp, Patel:2011th, Profumo:2014opa, Beniwal:2017eik, Beniwal:2018hyi,
Niemi:2018asa, Ramsey-Musolf:2019lsf, 
Hashino:2018zsi,Kozaczuk:2019pet, Niemi:2020hto, Chiang:2020rcv, Cline:2022xhx,Ellis:2022lft,
Chatterjee:2022pxf, Ghosh:2022fzp, Costa:2022lpy, Borah:2023zsb, Borah:2024emz, Kersten:2024ucm, YaserAyazi:2024dxk,
Jangid:2025ded, Mahapatra:2025vzu,
Bittar:2025lcr}. Other cosmological
and phenomenological consequences of an FOPT such as dark matter production, the generation of baryon asymmetry, 
and the formation of a primordial black hole (PBH), have been discussed in \cite{Baker:2019ndr, Chway:2019kft,  Jung:2021mku, 
Huang:2022vkf, Dasgupta:2022isg, Chun:2023ezg, Gouttenoire:2023naa, Ai:2024cka,Dutka:2024zwe,Bhandari:2025ufp,Murai:2025hse}.

The detection of the SGWB in the future GW detectors has also been investigated in the context of probing 
the seesaw scale \cite{Okada:2018xdh, Hasegawa:2019amx, Huang:2022vkf, Dasgupta:2022isg,Dichtl:2023xqd,Chun:2023ezg,
Datta:2024tne, Paul:2024iie, Barman:2025bir}.
In this work, we consider a gauged $\ubl$
extension of the SM with three heavy RHNs 
($\ni$, $i=1,\cdots, 3$) with $\rm B-L$ charge $-1$ and a complex scalar $\Phi$ 
which carries $\rm B-L$ charge $+2$. The former fields are required to satisfy the anomaly-free conditions 
whereas the latter field is required to break the $\ubl$ symmetry and generate the mass of the heavy RHNs. Thus, in 
this framework, one can dynamically generate the heavy mass of the RHNs once $\Phi$ acquires a nonzero vacuum expectation value (VEV). Here, we have studied
the production of baryon asymmetry from the decay of TeV scale RHNs. We would like to mention 
that our model is different from the one considered in Refs.\,\cite{Dasgupta:2022isg, Chun:2023ezg} where the authors have studied 
leptogenesis and an FOPT in scale invariant $B-L$ model. Because of the scale invariance, the predicted strength of the GW is different from that obtained in our study.
Additionally, in our analysis, we have considered (i) the flavor effects, which are important for such a low scale leptogenesis 
scenario, and (ii) the effect of the $\bl$ gauge boson $\zbl$ and $\Phi$ involving RHNs production channels. We find that the parameters of the $\ubl$ gauge sector 
as well as the Yukawa interaction strength between $\Phi$ and $\ni$
have strong dependence on the baryon asymmetry generation.

In addition, the scalar sector augmented with an extra scalar $\Phi$ can undergo an FOPT in the early Universe and produces an SGWB. Since 
the masses of the RHNs and $\zbl$ depend on the $\bl$ breaking scale, the SGWB production
also depends on the parameters of the $\ubl$ gauge sector. In light of this interconnection between
the production of baryon asymmetry and the SGWB from an FOPT, we investigate the possibility of probing the parameter space for leptogenesis 
in future GW detectors such as LISA \cite{Baker:2019nia}, 
BBO \cite{Corbin:2005ny, Crowder:2005nr, Harry:2006fi}, DECIGO, and Ultimate DECIGO (U-DECIGO) \cite{Seto:2001qf, Kawamura:2006up, Kawamura:2020pcg}. 
We show the region of model parameter space
where the observed baryon asymmetry can be explained via resonant flavored leptogenesis. Interestingly, a certain portion of this parameter space 
is capable of producing a detectable 
SGWB in future detectors. We also discuss the current collider constraints on our model parameters.

The rest of the paper is organized as follows. We briefly discuss our model in Sec. \ref{sec:model}.
The constraints on the model parameter space are discussed in Sec. \ref{sec:constraints}.
 A detailed discussion on resonant flavored leptogenesis is given in Section \ref{sec:leptogenesis}.
 Sec. \ref{sec:FOPT_GW} is devoted to a detailed discussion on FOPT dynamics, GW production from FOPT, and prospects
 of detecting SGWB in future GW detectors. Finally, we conclude in Sec. \ref{sec:summary}. The derivation of the $\cal A$
 matrix for flavored leptogenesis is given in Appendix \ref{AppA:A_matrix}. In Appendix \ref{appB:cross_sections}, all the 
 relevant cross sections for the analysis of resonant flavored leptogenesis are given.
 In Appendix \ref{appC:LFV}, we provide the explicit forms of the collision terms for a few lepton-number-violating processes. 
 The field-dependent and thermal masses of all the fields are given in Appendix \ref{AppD:masses}. 

\section{MODEL}
\label{sec:model}
In this section, we will describe our model briefly. We extend the SM gauge group $SU(3)_c \otimes SU(2)_L \otimes U(1)_Y$ with an additional $\ubl$ gauge symmetry. 
We introduce three RHNs $\ni$ (where $i=1,\,2,\,3$)  with $\rm B-L$ charge -1 in order to make the model anomaly-free. To break
the $\ubl$ gauge symmetry and generate
the mass of the RHN, a complex scalar $\Phi$ with $\rm B-L$ charge +2 has been introduced. 

The Lagrangian of the model is given by
\bal
\mathcal{L} = \mathcal{L}^{\rm SM}_{\rm Gauge} +  
\mathcal{L}^{\rm SM}_{\rm Yukawa} + \mathcal{L}_{\rm New}+ V(H, \Phi)\,\,,
\eal
where $\mathcal{L}^{\rm SM}_{\rm Gauge}$ and $\mathcal{L}^{\rm SM}_{\rm Yukawa}$ are the Lagrangians
for the SM gauge and Yukawa interactions, respectively. $V(H, \Phi)$ is the scalar potential, and it reads
\bal
\label{eq:scalar_pot}
V(H, \Phi) &= -\mu_H^2 H^\dagger H - \mu^2_\Phi \Phi^\dagger \Phi + \lambda_H (H^\dagger H)^2 + 
\lamf (\Phi^\dagger \Phi)^2 \nn\\
&-\lambda^\prime H^\dagger H \Phi^\dagger \Phi\,\,,
\eal
where $\mu_H$ and $\mu_\Phi$ are mass parameters of $H$ and $\Phi$, respectively and the quartic couplings of $H$ and 
$\Phi$ are denoted by $\lambda_H$ and $\lamf$ respectively. The explicit forms of
$H$ and $\Phi$ are given by
\bal
&H = \begin{pmatrix}
G^+\\
\dfrac{\zeta_1 + i \zeta_2}{\sqrt{2}}
\end{pmatrix}, ~~
\Phi = \dfrac{\eta_1 + i \eta_2}{\sqrt{2}}\,\,,
\eal
and at zero temperature, $\zeta_1$ and $\eta_1$  acquire nonzero VEV $\vh$ and $\vbl$, respectively. The last term of Eq.\,\eqref{eq:scalar_pot} 
is the mixing between the SM Higgs boson and $\ubl$ symmetry-breaking scalar, and it is proportional to the coupling $\lambda^\prime$. 
The parameters in Eq.\,\eqref{eq:scalar_pot} can be written in terms of the masses and mixing angle between scalars, as follows:
\allowdisplaybreaks
\bal
        & \mu_H^2 = \dfrac{(m_{\eta}^2 + m_{\zeta}^2) \vh - (m_{\eta}^2 - m_{\zeta}^2) (\vh \cos 2\theta + \vbl \sin 2\theta)}{4 \vh} \,\,,\nn\\
        & \mu_\Phi^2 = \dfrac{(m_{\eta}^2 + m_{\zeta}^2) \vbl + (m_{\eta}^2 - m_{\zeta}^2) (\vbl \cos 2\theta - \vh \sin 2\theta)}{4 \vbl}\,\,,\nn\\
        & \lambda_H = \dfrac{m_{\eta}^2 + m_{\zeta}^2 - (m_{\eta}^2 - m_{\zeta}^2) \cos 2\theta}{4 \vh^2}\,\,,\nn\\
       & \lambda_\Phi = \dfrac{m_{\eta}^2 + m_{\zeta}^2 + (m_{\eta}^2 - m_{\zeta}^2) \cos 2\theta}{4 \vbl^2}\,\,,\nn\\
        & \lambda^\prime = -\dfrac{(m_{\zeta}^2 - m_{\eta}^2) \sin \theta \cos \theta}{\vh \vbl}\,\,\,,
\eal
where $\eta$ and $\zeta$ are the two scalar degrees of freedom in the mass basis with masses $m_{\eta}$ and $m_{\zeta}$ respectively. In the limit $\theta \to 0$,
$m_{\zeta}^2 \to 2 \lambda_H \vh^2$, and thus $\zeta$ is identified as an SM Higgs boson.

The interactions of the SM fields with a $\rm B-L$ gauge boson
$\zbl$, $\Phi$ as well as the kinetic terms of $\zbl$, $\Phi$, and $\ni$ are contained in $\mathcal{L}_{\rm New}$,
and it can be written as
\bal
\mathcal{L}_{\rm New} &= 
-\dfrac{1}{4} X^{\mu \nu}X_{\mu \nu} + 
\sum_{i=1}^3
i\overline{\ni}_R \slashed{D} {\ni}_R \nn\\
&-\sum_{i=1}^3\left(\dfrac{Y_i}{2}
\overline{N_{i_R}^c} N_{i_R}\Phi + {\rm H.c} 
\right) -\gbl \sum_{f \ne \nu} Q^{(f)}_{\rm BL}\bar{f} \gamma_\mu f \zbl^\mu \nn\\
&-\gbl \sum^3_{i=1} Q^{(\nu)}_{\rm BL}\bar{\nu}_i \gamma_\mu P_L \nu_i \zbl^\mu -
\sum_{\alpha,i} {Y_{D}}_{\alpha i} 
\overline{L_{\alpha}} \Tilde{H} N_{iR}\,\,,
\label{eq:Lag_BL}
\eal
where $f$ is the charged SM fermion, and $\nu_i$
 ($i=1,\,2,\,3$) is the SM neutrino. $Q^{(f)}_{\rm BL}$ and $Q^{(\nu)}_{\rm BL}$ are the $\bl$ charges
 of the charged SM fermions and SM neutrinos respectively. 
 $Q^{(f)}_{\rm BL} = -1 \,\,(1/3)$ for the charged SM leptons (quarks)
 and $Q^{(\nu)}_{\rm BL} = -1$ for the SM neutrinos. The $\ubl$
 gauge coupling is denoted by $\gbl$, and 
 $X^{\mu \nu} = \partial^\mu \zbl^\nu -\partial^\nu \zbl^\mu$ is the field strength tensor of the $\rm B-L$ gauge boson $\zbl$.
 In the present scenario, we have not considered
 kinetic mixing between $\ubl$ and $U(1)_{\rm Y}$.
 This assumption can be justified because introducing
 kinetic mixing does not reveal any new physics
 since the SM fermions are already coupled to $\zbl$
 due to their nonzero $\bl$ charge. Therefore,
 the effect of $Z-Z_{\rm BL}$ mixing on top of the
 tree-level interaction can be safely neglected.
 The Yukawa coupling coefficient between $\ni$ and
 $\Phi$ is denoted by $Y_i$. The last term is the
 Dirac-Yukawa interaction among the lepton doublet,
 the Higgs doublet, and the RHN. This interaction is
 responsible for the Dirac mass of the neutrinos, similar
 to the other charged fermions in the SM. Moreover,
 the lepton asymmetry generated from the out-of-equilibrium
 decay of RHNs is also due to the Dirac-Yukawa interaction.
 In the present scenario, we have used the Casas-Ibarra
 parametrization \cite{Casas:2001sr} to calculate the
 Dirac-Yukawa coupling matrix $Y_{D}$. According to
 this formalism, we can express $Y_D$ as
 \begin{eqnarray}
 Y_D = M^{1/2}\,R\,D^{1/2}_{\nu}\,U^\dagger \,, 
 \label{eq:Casas-Ibarra}
 \end{eqnarray}
where $M$, $D_\nu$ are the RHN and light neutrino 
mass matrices, respectively, and they are assumed
to be diagonal. The matrix $U$ is the Pontecorvo–Maki–Nakagawa–Sakata 
matrix and $R$ is an orthogonal matrix with
complex parameters. In this work,
we take a simple form of $R$ as
\begin{eqnarray}
R =   
\begin{pmatrix}
0 & \cos{\Theta} & \sin{\Theta} \\
0 & -\sin{\Theta} & \cos{\Theta}\\
1 & 0 & 0
\end{pmatrix} ,
\label{eq:Rmatrix}
\end{eqnarray}
with $\Theta = \theta_R + i \, \theta_I$,
and the explicit values of $\theta_R$, and $\theta_I$
will be discussed in the next section. The particular form
of $R$ ensures that the Yukawa coupling matrix $Y_D$ is independent
of $M_3$ when the lightest active neutrino mass is chosen 
to be zero, i.e., $m_1 = 0$ \cite{Granelli:2020ysj}.
Moreover, in this formalism,
the Yukawa couplings (${Y_{\rm D}}_{3i}$, $i=1,\,2,\,3$)
associated with the third generation of the SM leptons
are zero. As a result, no {\it CP} asymmetry is generated
for the third generation from the decays of RHNs.

The $\ubl$ symmetry is spontaneously broken when the scalar $\Phi$ gets a VEV, and as a result the $\bl$ 
gauge boson and the RHNs become massive. Therefore, after symmetry breaking the masses of $Z_{\rm BL}$
and $\ni$s are given by
\begin{eqnarray}
m_{\zbl} &=& 2\, \gbl v_{\rm BL}\,, \\
M_{i} &=& Y_i \dfrac{v_{\rm BL}}{\sqrt{2}}\,.
\label{eq:B_L_masses}
\end{eqnarray}
We will see in Sec.\,\ref{sec:leptogenesis} that for successful low scale leptogenesis,  we require the lightest two RHNs to be
quasidegenerate i.e., $Y_1 \approx Y_2$, while the 
mass gaps between $N_3$, and $N_{1,2}$ are much larger than the decay widths of $N_{1,2}$, so that the asymmetry
generated from 
$N_3$ is not resonantly enhanced. In our analysis, we have considered 
$Y_3 \gg Y_1,\,Y_2$
so that $M_3 \gg M_1,\,M_2$. 
The choice of this mass hierarchy implies that the lepton asymmetry is produced only from the decays of $N_1$, and $N_2$,
and at the time of asymmetry generation, the abundance of $N_3$
is Boltzmann suppressed.

\section{RELEVANT CONSTRAINTS}
\label{sec:constraints}
\subsection{Constraint from electroweak precision observables (EWPO)} 
The new physics effect on the EWPO observables can be
    parametrized by the oblique parameters $S$, $T$, and $U$ \cite{Peskin:1991sw}. In our scenario, the self-energy correction of $ZZ$ and $W^+W^-$
    gets an additional contribution due to the mixing between $\Phi$ and the Higss field, and as a result, the oblique parameters
    are modified. The change in the oblique parameters in the presence of a new scalar field is given by
    \bal
    \Delta T &= \dfrac{3}{16 \pi \sin^2 \theta_W} \left[
    \cos^2 \theta \left\{f_T\left(\dfrac{m_{\zeta}^2}{m_W^2}\right) - 
    \dfrac{1}{\cos^2 \theta_W^2} f_T\left(\dfrac{m_{\zeta}^2}{m_Z^2}\right)\right\} \right.\nn\\
    &\left.+\sin^2 \theta \left\{f_T\left(\dfrac{m_{\eta}^2}{m_W^2}\right) - \dfrac{1}{\cos^2 \theta_W^2} f_T\left(\dfrac{m_{\eta}^2}{m_Z^2}\right)\right\} \right.\nn\\
    &\left.-\left\{f_T\left(\dfrac{m_{\zeta}^2}{m_W^2}\right) - \dfrac{1}{\cos^2 \theta_W^2} f_T\left(\dfrac{m_{\zeta}^2}{m_Z^2}\right)\right\}
    \right]\,\,,\nn\\
    \Delta S &= \dfrac{1}{2\pi} \left[\cos^2 \theta \,f_S \left(\dfrac{m_{\zeta}^2}{m_Z^2}\right) + \sin^2 \theta\, f_S \left(\dfrac{m_{\eta}^2}{m_Z^2}\right)
    -f_S \left(\dfrac{m_{\zeta}^2}{m_Z^2}\right)\right]\,\,,\nn\\
    \Delta U &= \dfrac{1}{2\pi} \left[\cos^2 \theta \,f_S \left(\dfrac{m_{\zeta}^2}{m_W^2}\right) + \sin^2 \theta\, f_S \left(\dfrac{m_{\eta}^2}{m_W^2}\right) 
    -f_S \left(\dfrac{m_{\zeta}^2}{m_W^2}\right)\right]\nn\\
    &-\Delta S\,\,,
    \eal
    where $\theta_W$ is the Weinberg angle, $\sin^2 \theta_W = 0.231$, and  $m_{\zeta}= 125 \,\rm GeV$ \cite{ParticleDataGroup:2024cfk}. 
    The loop functions $f_T(x)$ and $f_S(x)$ are given by
   \cite{Baek:2011aa}
    \bal
    &f_T(x) =  \dfrac{x \log x}{x-1}\,\,,\nn\\
    &\underline{\text{ for } 0<x<4: } \nn\\
    &f_S(x) = \dfrac{1}{12}\left[-2 x^2 + 9x + \left\{(x-3)(x^2 - 4x + 12) \right.\right.\nn\\
    &\left.\left.+ \dfrac{1-x}{x}\right\} f_T(x) 
   +\sqrt{x(4-x)} (x^2 - 4x + 12) \tan^{-1} \left(\sqrt{\dfrac{4-x}{x}}\right)
    \right]\,,\nn\\
    &\underline{\text{ for } x\ge 4: } \nn\\
    & f_S(x)= \dfrac{1}{12}\left[-2 x^2 + 9x + \left\{(x-3)(x^2 - 4x + 12) \right.\right.\nn\\
    &\left.\left.+ \dfrac{1-x}{x}\right\} f_T(x)  
    +\sqrt{x(x-4)} (x^2 - 4x + 12) \right.\nn\\
    &\left.\times\log \left(\dfrac{x - \sqrt{x(x-4)}}{x + \sqrt{x(x-4)}}\right)
    \right]\,.
    \eal

    To derive the constraint from the EWPO, we use $\Delta S = 0.04 \pm 0.11$, $\Delta T = 0.09 \pm 0.14$, and $\Delta U = -0.02 \pm 0.11$ and 
    following \cite{Profumo:2014opa, Beniwal:2018hyi}, we define $\Delta \chi^2$ as
    \bal
    \Delta \chi^2 = \sum_{i, j} \left(\Delta \mathcal{O}_i - \Delta {\mathcal{O}_{{\rm obs}, i}}\right) \left[\mathcal{C}^{-1}\right]_{ij}
    \left(\Delta \mathcal{O}_j - \Delta{\mathcal{O}_{{\rm obs}, j}}\right),\nn
    \eal
    where
    \bal
    \Delta \mathcal{O} = \left(\Delta S\,\,\Delta T\,\,\Delta U\right)^T, ~
    \Delta \mathcal{O}_{\rm obs} = \left(0.04\,\,0.09\,\,-0.02\right)^T.
    \nn
    \eal
    The $ij$ element of the covariance matrix $\mathcal{C}$ is $\sigma_i \rho_{ij} \sigma_j$,
    and the explicit forms of $\sigma$ and $\rho$
    are given by 
    \bal
    &\sigma = \left(0.11~~~0.14~~~0.11\right)^T\,,\nn\\ 
    &\rho = \begin{pmatrix}
    1 & 0.92 & -068 \\
    0.92 & 1 & -0.87 \\
    -0.68 & -0.87 & 1
    \end{pmatrix}\,\,.
    \eal
    We determine the allowed model parameters at $95\%$ confidence interval by imposing the condition $\Delta \chi^2 < 5.99$.

  \subsection{Higgs signal strength} 
  The mixing angle between the scalar fields can be constrained from the measurement of
    the Higgs signal strength at the LHC. Combining the constraints in different channels, the current bound 
    on the Higgs signal strength ($r_h$) is $1.02 ^ {+0.11}_{-0.12}$, which reflects the absence of any deviation from the SM prediction \cite{Ellis:2013lra}.
    The signal strength $r_h$ in any channel $XX$ is defined as \cite{Li:2014wia}
    \bal
    r_h &= \dfrac{\sigma_{gg \to \zeta} {\rm Br}(\zeta \to XX)}{\sigma_{gg \to h}^{\rm SM} {\rm Br}(h \to XX)_{\rm SM}}\nn\\
    & = \cos^2 \theta \times \cos^2 \theta \times \dfrac{\Gamma_h^{\rm SM}}{\Gamma_{\zeta}}\,\,,
    \eal
    where $\Gamma_h^{\rm SM} = 3.7\,\rm MeV$ is the total decay width of the SM Higgs boson \cite{ParticleDataGroup:2024cfk}. 
    $\Gamma_{\zeta} = \cos^2 \theta \,\Gamma^{\rm SM}_h + \sum_i \Gamma_{\zeta \to \ni \ni} \Theta (m_{\zeta} - 2 M_{i}) + \Gamma_{\zeta \to \eta \eta}
    \Theta (m_{\zeta} - 2 m_{\eta})$. In our parameter space of interest, $\Gamma_{\zeta \to \ni \ni} = \Gamma_{\zeta \to \eta \eta}=0$
    and the Higgs signal strength bound can be translated into the bound on the mixing angle 
    as $\sin^2 \theta \le 0.22$ at $95\%$ CL.
 
 \subsection{LEP and LHC constraint on $\boldsymbol{\zbl}$} 
        The measurement of $e^+e^- \to \ell^+ \ell^-$ at the LEP constrains the $m_{\zbl}-\gbl$ plane as $m_{\zbl}/\gbl > 6\,\rm TeV$ \cite{Escudero:2018fwn}.
        However, for $m_{\zbl} < 5\,\rm TeV$, the bound from the dilepton search at the LHC is stronger than the LEP constraint \cite{FileviezPerez:2020cgn}. 
\section{FLAVORED LEPTOGENESIS}
\label{sec:leptogenesis}
The out-of-equilibrium decays of RHNs into the SM lepton and scalar can generate an asymmetry between 
the particle and antiparticle in the leptonic sector if there are some lepton-number-violating interactions and 
simultaneous {\it C} and {\it CP} violations as well. The lepton asymmetry thus generated is efficiently converted into the 
baryon asymmetry by the $\bl$ conserving sphaleron processes as long as the sphalerons are
in thermal equilibrium with the SM plasma ($T\gtrsim 130$ GeV). In the present scenario, 
we consider the masses of the RHNs and $\zbl$
are of the order of TeV scale to test this model in future GW detectors, as discussed in 
Sec.\,\ref{sec:FOPT_GW}, and this can be achieved by choosing $\vbl$, $\gbl$, and $\yf$ 
appropriately. This restriction automatically
implies that the masses of the RHNs, generated through the spontaneous
breaking of $U(1)_{\bl}$ symmetry would be much smaller than
the canonical mass scale of the RHN ($\gtrsim 10^{9}$ GeV) 
required for the vanilla leptogenesis scenario. Therefore, 
the leptogenesis occurs at a much lower temperature,
and in this case, the flavor effect becomes important. The
main reason is that when the temperature is above $10^{12}$ GeV,
we do not need to consider three lepton flavor states separately,
instead they can be considered as a coherent superposition
of all three flavor states ($e,\,\mu$, and $\tau$). However, the coherence is lost due to the Yukawa interaction which becomes
effective at different temperatures for different flavors. Consequently, we have three 
distinct regimes namely, the unflavored regime, partly flavored regime, and fully flavored regime, respectively. 
The unflavored regime refers to an era when $T\gtrsim 10^{12}$ GeV. When the temperature drops below $10^{12}$ GeV, 
the Yukawa interaction of the $\tau$ lepton becomes larger than the corresponding Hubble expansion rate, and the coherence between 
$\tau$ and the other two flavors ($e$ and $\mu$) is lost. This is
known as a partly flavored regime since in this regime, one needs to consider the $\tau$ lepton separately while $e$ 
and $\mu$ still maintain coherence. This situation continues until the temperature
drops below $10^{9}$ GeV when the Yukawa interaction of the muon enters equilibrium, and 
all three flavors now become distinctly different. This is known as the fully flavored regime. As mentioned 
above, the mass scale of the RHNs is a few TeV; therefore we are essentially working in the fully flavored regime 
where lepton asymmetry for all three flavors needs to be computed separately by solving individual Boltzmann 
equations. Moreover, at this low energy scale $T\sim M_1$ ($M_1$ is the mass of the lightest RHN), as the desired 
lepton asymmetry cannot be generated through vanilla leptogenesis, we need some mechanism to enhance the {\it CP} asymmetry. 
This can be achieved due to a resonant enhancement in the {\it CP} asymmetry parameter if we have two quasidegenerate RHNs. 
The resonant leptogenesis is immensely popular from the point of view of the detection prospect of the RHN states in 
ongoing experiments, particularly at the LHC, which is capable of producing TeV scale particles from $pp$ collisions \cite{Dev:2013wba, Deppisch:2015qwa}. 

Let us consider $Y_{\ell_{L_{\alpha}}}$ as the coming number density of the lepton doublet of flavor $\alpha$, 
which is defined as $ Y_{\ell_{L_{\alpha}}} = n_{\ell_{L_{\alpha}}}/s$ where $n_{\ell_{L_{\alpha}}}$
is the number density of flavor $\alpha$, and $s$ is the entropy density.
Now, we define the comoving number density for the lepton asymmetry of flavor
$\alpha$ as $\Delta Y_{\ell_{L_{\alpha}}} = Y_{\ell_{L_\alpha}} - Y_{\bar{\ell}_{L_\alpha}}$. However, $\Delta Y_{\ell_{L_\alpha}}$ is not a conserved quantity 
since in the SM, the lepton number is violated by the sphaleron processes. Thus, we would like to calculate $\Delta B/3- \Delta L_{\alpha}$ which is conserved by the sphaleron processes and $L_\alpha$ is the total lepton number of flavor $\alpha$. Therefore, $\Delta Y_{\ell_{L_\alpha}}$, generated through the decays of the RHNs is redistributed to the the $B/3-L_\alpha$ asymmetry
via $B/3-L_\alpha$ conserving sphaleron processes, and it can be expressed as
\begin{eqnarray}
\Delta Y_{{\ell_{L_\alpha}}} = -\sum_{\beta = e,\,\mu,\,\tau}
\mathcal{A}_{\alpha \beta}\,\,Y_{\Delta_{\beta}} \,,
\label{eq:A_matrix}
\end{eqnarray}
where $\Delta_{\alpha} \equiv \Delta{B}/3 - \Delta{L_{\alpha}}$, and the
$\mathcal{A}$ matrix 
when all the SM Yukawa couplings are in equilibrium 
(when $T \le10^{5}$ GeV) is given by \cite{Jones:2013czt}
\begin{eqnarray}
\mathcal{A} = 
\dfrac{1}{711}\begin{pmatrix}
442 & -32 & -32 \\
-32 & 442 & -32 \\
-32 & -32 & 442
\end{pmatrix}\,.
\end{eqnarray}
It is a symmetric matrix with $\mathcal{O}(1)$ diagonal
elements and tiny off-diagonal elements. 
The baryon
asymmetry $\Delta Y_B$
is related to $Y_{\Delta_{\alpha}}$
by the sphaleron factor
$28/79$ as
\begin{eqnarray}
\Delta Y_{B} 
&=&\dfrac{28}{79} \sum_{\alpha} Y_{\Delta_{\alpha}} \,.
\label{eq:baryon_asymmetry}
\end{eqnarray}
The derivation 
of the $\cal A$ matrix and the sphaleron factor 
is given in Appendix \ref{AppA:A_matrix}.

Finally, $\Delta Y_{B}$ is related to the
baryon-to-photon ratio of the Universe as
\begin{eqnarray}
\eta_{B} &=& \dfrac{\Delta Y_B\,s}{n_{\gamma}} \,,\\
&=& 7.0397\,\Delta Y_{B}\,, 
\label{eq:etaB}
\end{eqnarray}
where $n_{\gamma} = 410.7 {\rm cm}^{-3}$ and
$s = 2891.2 {\rm cm}^{-3}$ are the number density of the 
photon and the entropy density of the Universe at the
present epoch respectively. The current value of the baryon-to-photon ratio
is $\eta_{B} = (6.143 \pm 0.190)\times 10^{-10}$ \cite{ParticleDataGroup:2024cfk}. 

Keeping in mind the above discussions, we will now write the Boltzmann equations
for the comoving number densities of the RHNs and $\Delta_{\alpha}$,
respectively,
\bal
&\dfrac{d Y_{\ni}}{dz} = -\dfrac{1}{z H}
\left[
\langle \Gamma_i \rangle \left(Y_{\ni} - Y^{\rm eq}_{\ni} \right) \right.\nn\\
&\left.-
\langle \Gamma_{\zbl \to \ni \ni} \rangle\,Y^{\rm eq}_{Z_{\rm BL}}
\left(1 - \left(\dfrac{Y_{\ni}}{Y^{\rm eq}_{\ni}}\right)^2 \right)
+ \dfrac{\gamma_{\rm tot}}{s} \left(
\left(\dfrac{Y_{\ni}}{Y^{\rm eq}_{\ni}}\right)^2-1\right) \right.\nn\\
&\left. + \dfrac{\gamma_{\Delta \ell = 1}}{s}\left(
\dfrac{Y_{\ni}}{Y^{\rm eq}_{\ni}} - 1\right)
\right]\,\,,
\label{eq:BE_N1}
\eal
where $z = M_1/T$, and $H$ is the Hubble parameter. As mentioned above, here we are
considering two quasidegenerate RHNs to obtain a significant
resonant enhancement in the {\it CP} asymmetry parameter. The remaining RHN ($N_3$)
is chosen to have a large mass gap from these two quasi-degenerate
RHNs (${N}_1$ and ${N}_2$) such that the effect of
$N_3$ on the lepton asymmetry
can be neglected. In the above, $Y_i$ 
and $Y^{\rm eq}_{\ni}$
($i=1,\,2$) are the comoving number density
and equilibrium comoving number density
of $\ni$, respectively. The total decay width of $\ni$ is $\Gamma_i = \sum_\alpha \left(\Gamma_{\ni \to \ell_\alpha H} + 
\Gamma_{\ni \to \bar{\ell}_\alpha \bar{H}}\right)$,  and
the thermal averaged decay width
$\langle \Gamma_ i\rangle = \Gamma_i {\rm K}_1(z)/{\rm K_2}(z)$.
The explicit form of $\Gamma_i$ is given by
\bal
\label{eq:RHN_decay_width}
\Gamma_i = \dfrac{\left(\Yd^\dagger \Yd\right)_{ii}}{16\pi}\,M_i\,,
\eal
where $i$ is not summed over. The second term represents the
change in the comoving number density of $\ni$ due to $Z_{\rm BL}$
decaying into a pair of $\ni$, and the corresponding decay width
is denoted by $\Gamma^{ii}_{\rm BL}$ with $Y^{\rm eq}_{Z_{\rm BL}}$
being the equilibrium comoving number density of the $\bl$ gauge boson.

Apart from the decay, the abundance of $\ni$ depends on
the third-generation quarks\footnote{Here we have ignored the effect of the first- and second-generation quarks due to tiny Yukawa couplings.} 
involving the $\Delta L_\alpha=1$ scattering processes and the effect of these processes in the evolution of $\ni$ will appear 
through the quantity $\gamma_{\Delta L_\alpha = 1}$ and it is defined as
\bal
\label{eq:gamma_delta_1}
\gamma_{\Delta \ell = 1} = \gamma^{\bar{t}_R U_{3_L}}_{\ell_{\alpha_L} \ni} + 
\gamma^{\ell_{\alpha_L} t_R}_{U_{3_L} \ni} + 
\gamma^{\bar{\ell}_{\alpha_L} U_{3_L}}_{ t_R \ni}\,\,.
\eal
The general form of $\gamma^{ij}_{kl}$ for the scattering $ij \rightarrow kl$ is given by 
\bal
\label{eq:gamma_ijkl}
\gamma^{ij}_{kl} = M_1^5\dfrac{2\pi^2 T}{(2 \pi)^6}
 \int_{4}^\infty dx \sqrt{x}\,{\rm K}_1 \left(\sqrt{x} z\right)
 \bar{\sigma}_{ij \to kl}\,\,,
\eal
where $\bar{\sigma}_{ij \to kl} \equiv 2\, g_i g_j 
x\,\Lambda\left(1, \frac{r_i}{x}, \frac{r_j}{x} \right)
\sigma_{ij \to k l}$ with $\Lambda(a,b,c) = (a-b-c)^2 - 4bc$
and $r_i = \frac{M^2_i}{M^2_1}$.
$g_i$($g_j$) and $M_i$($M_j$) are the internal degrees of
freedom and the mass of the initial state particle $i$($j$) respectively, 
and $\sigma_{ij \to kl}$ is the cross section of the $ij \to kl$ process.

Additionally, the abundance of $\ni$ strongly depends on $\gbl$ and $\yf$ through $\zbl$ and $\eta$ involving the scattering process, 
and the information of these processes is encoded in the quantity 
$\gamma_{\rm tot}$  which is given by
\bal
\label{eq:new_scattering_gamma}
\gamma _{\rm tot} = 
\gamma^{\ni \ni}_{\zbl \zbl} + \gamma^{\ni \ni}_{\eta \eta} + \gamma^{\ni \ni}_{\zbl \eta}\,\,.
\eal
The explicit forms of $\bar{\sigma}_{ij \to kl} $ for the different processes mentioned in Eqs.\,\eqref{eq:gamma_delta_1} and \eqref{eq:new_scattering_gamma} 
are given in  Appendix \ref{appB:cross_sections}. We would
like to note here that in deriving the Boltzmann equation of $\ni$, we have assumed
that BSM degrees of freedom, except the RHNs such as $\zbl$ and $\eta$, are in thermal equilibrium with the SM plasma. This can be achieved easily if we
have either a sufficiently large gauge coupling $\gbl$ or a large
scalar mixing angle $\theta$ with the Higgs boson or both.
\begin{figure}[h]
\centering
\includegraphics[width = 0.45\textwidth]{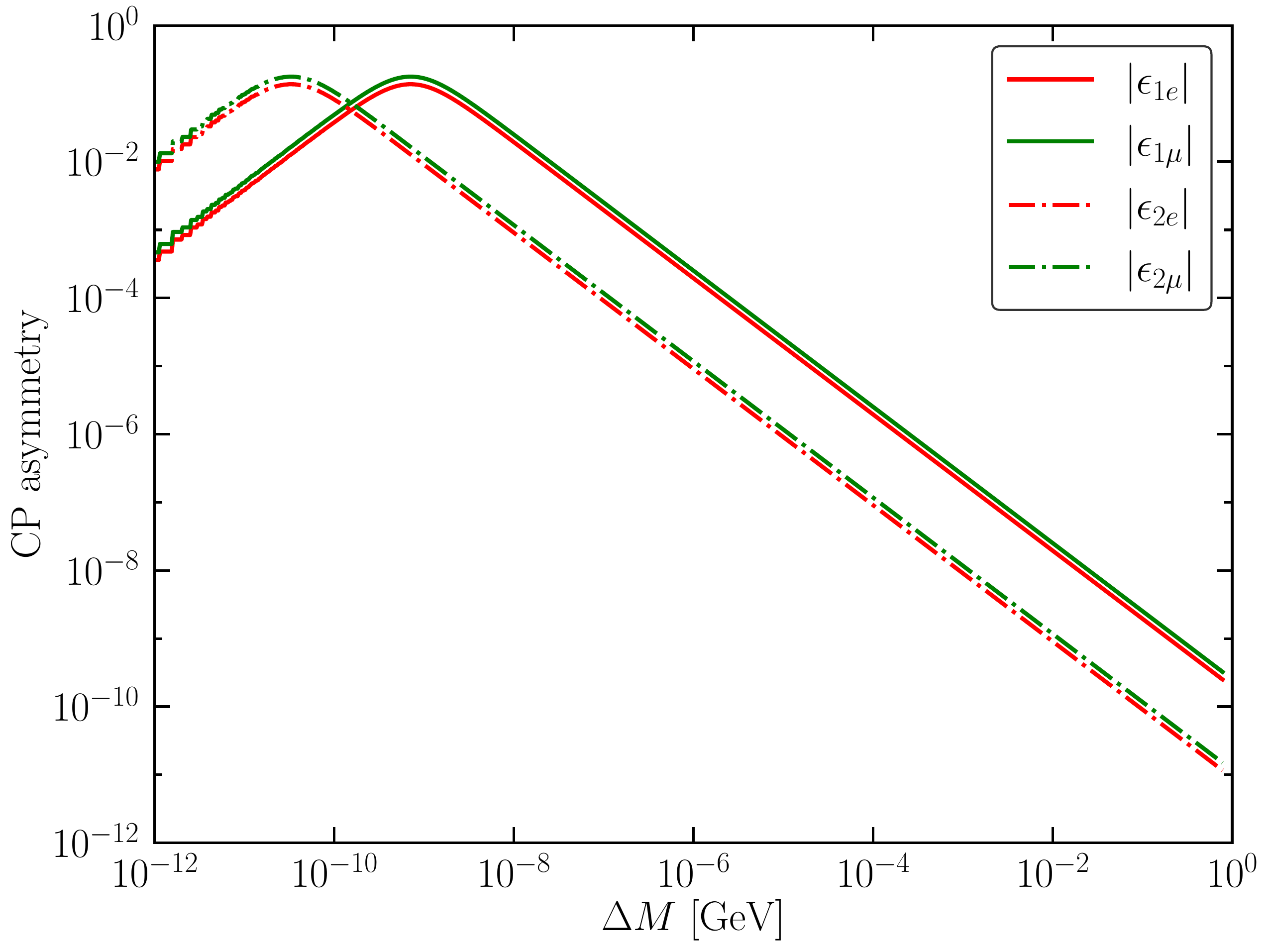}
\caption{Variation of {\it CP} asymmetry with mass splitting $\Delta{M}$
for $M_1 = 3$ TeV.}
\label{fig:CP_asymmetry}
\end{figure}
The collision term of the Boltzmann equation for $Y_{\Delta_{\alpha}}$
is composed of contributions coming from the decay and inverse decay of
$\ni$'s and other lepton-number-violating scatterings. The Boltzmann equation for the lepton flavor $\alpha$ can be written as
\begin{figure}[hbt!]
\centering
\includegraphics[width = 0.45\textwidth]{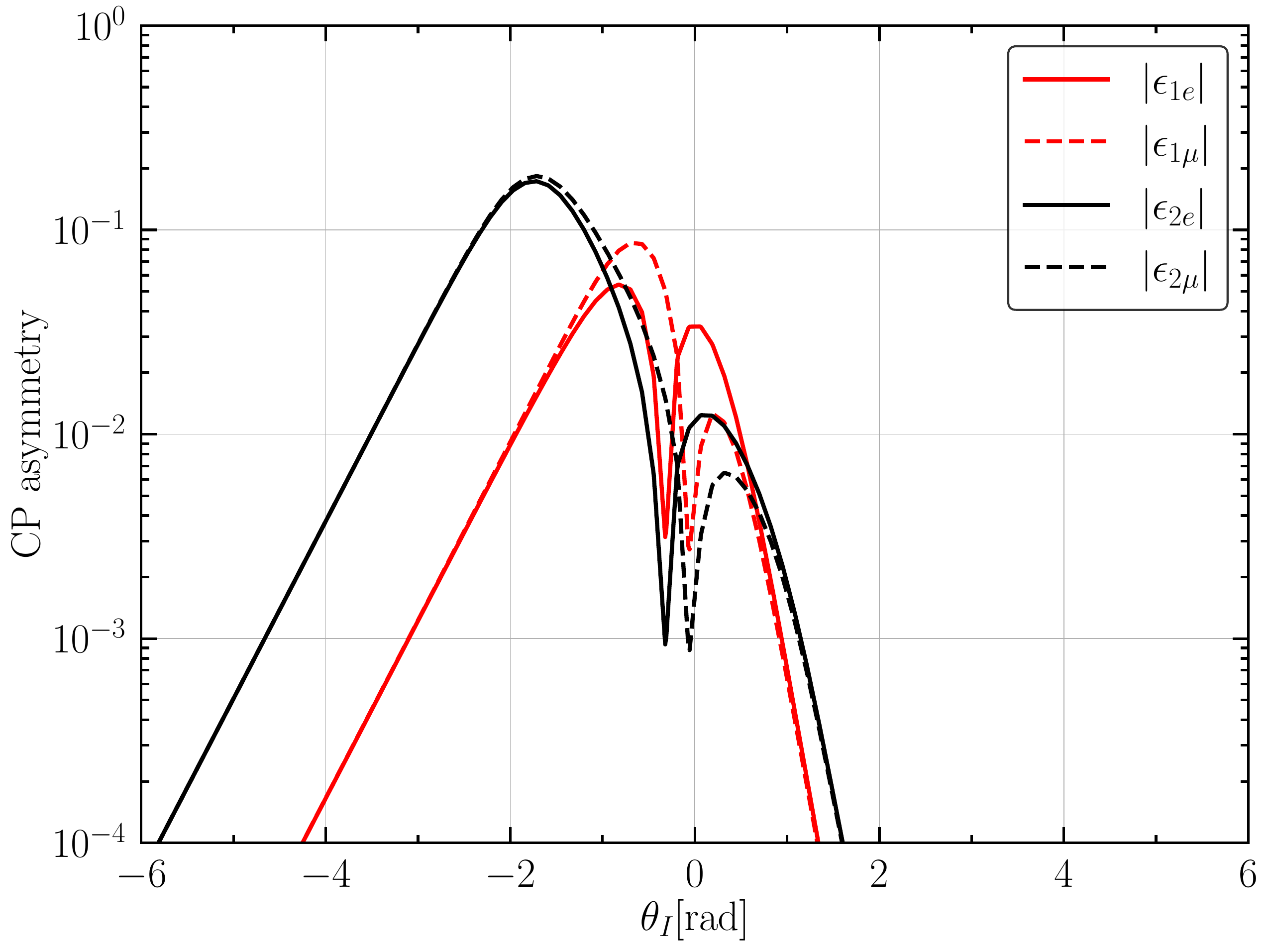}
\caption{Variation of {\it CP} asymmetry with $\theta_I$
for $M_1 = 3$ TeV,
$\theta_R = 1\,\rm rad$, and
$\Delta M = 1.3 \times 10^{-10}$ GeV.}
\label{fig:CP_asymmetry_vs_thetaI}
\end{figure}
\begin{figure}[t!]
\centering
\includegraphics[width = 0.45\textwidth]{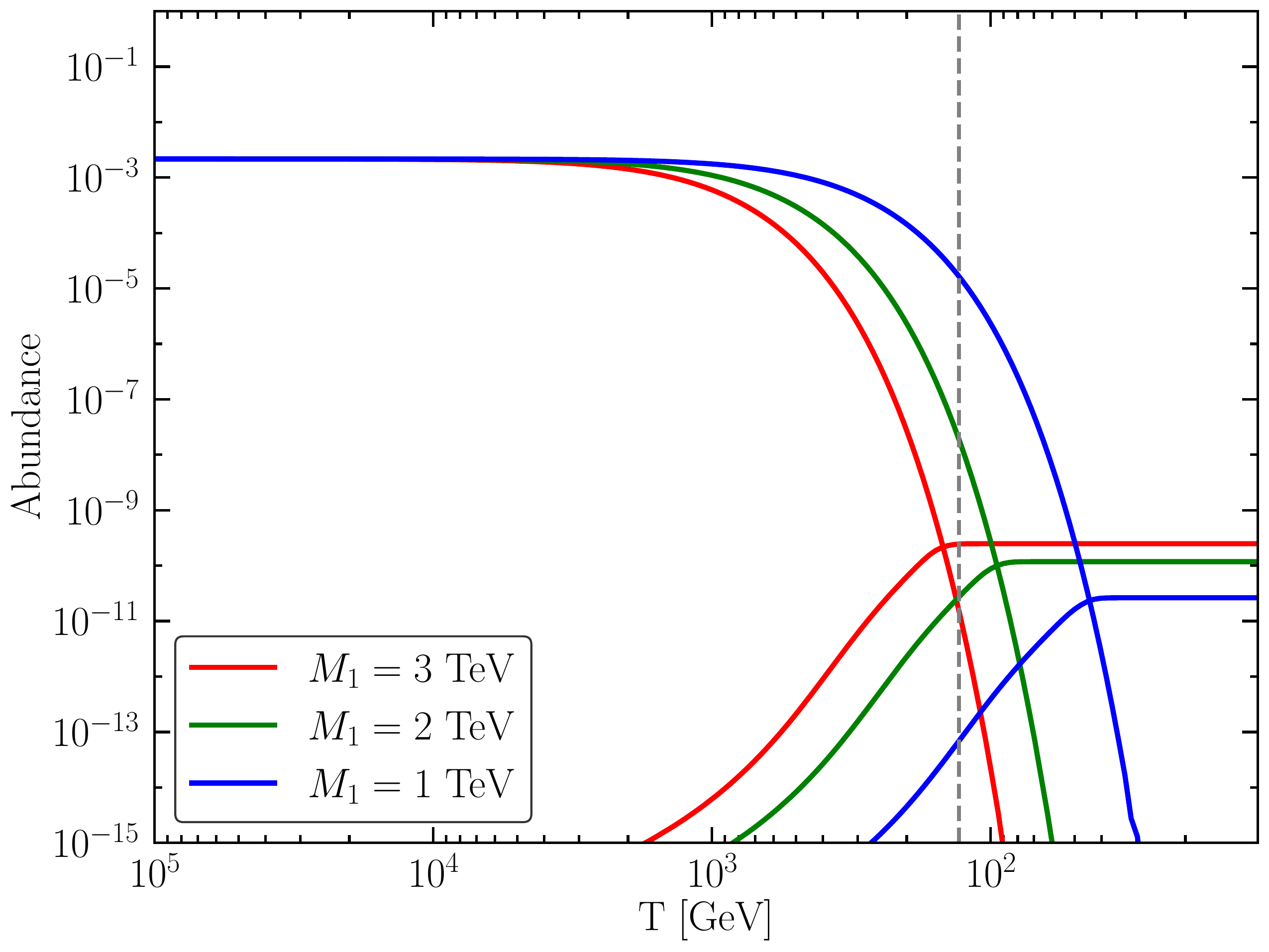}
\caption{Evolution of $Y_\Delta \equiv Y_{{{\Delta}_e}} + Y_{{{\Delta}_\mu}}$ as a function of $T$ for
$M_1 = 1$ TeV (blue), $2$ TeV (green) and $3$ TeV (red), respectively. The vertical dashed line denotes
the sphaleron decoupling temperature.}
\label{fig:comparison_plot}
\end{figure}
\begin{widetext}
\bal
\label{eq:lepton_assymetry_BE}
\dfrac{d Y_{\Delta_{\alpha}}}{dz} &= 
-\dfrac{1}{z H}
\sum_i
\left[
\langle \epsilon _{i \alpha} \Gamma_i\rangle
\left(Y_{\ni} - Y^{\rm eq}_{\ni}  \right)
-\dfrac{\Delta Y_{{\ell}_{L_\alpha}}}{2 Y_{\ell_\alpha}^{\rm eq}}\,
\langle \Gamma^{\alpha}_i \rangle\,Y^{{\rm eq}}_{\ni} 
-\dfrac{1}{s}
\dfrac{\Delta Y_{\ell_{L_\alpha}}}{Y_{\ell_\alpha}^{\rm eq}}
\left(
 \dfrac{Y_{\ni}}{Y^{\rm eq}_{\ni}}
\gamma^{\bar{ t}_R {U_3}_L}_{i \alpha}+ 
\gamma^{\alpha t_R}_{i {U_3}_L}
+ 
\gamma^{\bar{\alpha} {U_3}_L}_{i t_R}
\right) + \mathcal{C}_{\rm LFV}
\right]\,\,,
\eal
\end{widetext}
where in the right-hand side, we replace $\Delta Y_{\ell_{L_\alpha}}$
with Eq.\,\eqref{eq:A_matrix}.

Here, the first term is the creation of
asymmetry in $\alpha$ flavor due to the decay of $\ni$ into $\ell_{L_\alpha}$ and the corresponding
scalar, while the second term represents
a washout of the created lepton asymmetry as a result of the inverse decay. The terms within the parentheses are the 
contributions coming from the $\Delta{L_{\alpha}} = 1$ scattering processes in the evolution of the lepton asymmetry. 
The last ${\cal C}_{\rm LFV}$ term contains subdominant lepton number violating processes and the explicit form of the collision terms are given in Appendix \ref{appC:LFV}. 

The asymmetry generated from the decay of $\ni$ into a particular flavor $\alpha$ is parametrized by a parameter called the {\it CP}
asymmetry parameter, which is defined as
\bal
\epsilon_{i \alpha} = \dfrac{\Gamma_{\ni \to \ell_\alpha H} -\Gamma_{\ni \to \bar{\ell}_\alpha \bar{H}}}
{\Gamma_i}\,\,,
\eal
where the summation over $SU(2)_L$ indices is implicit. The explicit form of $\epsilon_{i\alpha}$ in terms of the model parameters is as follows:
\bal
\label{eq:asymmetry_para}
\epsilon_{i \alpha} = \dfrac{1}{8\pi \left(\Yd^\dagger \Yd\right)_{ii}}
\sum_{j \ne i}&\left[
\sqrt{x_{ji}} 
\left(1 + (1 + x_{ji}) \log \dfrac{x_{ji}}{1 + x_{ji}}
\right) \kappa_1\, 
+ \right.\nn\\
&\left. + 
\dfrac{(1-x_{ji})}
{\left(1-x_{ji}\right)^2+ x_{ji} {\cal Y}_{ji}} \kappa_2
\right]\,,
\eal
where
\bal
\kappa_1 &= {\rm Im} \left[{\Yd}^*_{\alpha i} {\Yd}_{\alpha j}
({\Yd}^T {\Yd}^*)_{ji}
\right]\,, \nn\\
\kappa_2  &= \,{\rm Im} \left[{\Yd}^*_{\alpha i} {\Yd}_{\alpha j} ({\Yd}^\dagger {\Yd})_{ji} \right.\nn\\
&\left.+{\Yd}^*_{\alpha i} {\Yd}_{\alpha j} ({\Yd}^T {\Yd}^*)_{ji}\,\sqrt{x_{ji}}
\right]\,.\nn
\eal
Here, ${\cal Y}_{ji} = \Gamma_j^2/M_i^2$,
$x_{ji} = M_j^2/M_i^2$, and the index $i$ is not summed over.
At tree level, there is no {\it CP} asymmetry between the two decay modes
($\ni \rightarrow \ell_{L_\alpha} H$ and $\ni \rightarrow \bar{\ell}_{L_\alpha} \bar{H}$). However, an asymmetry can be
generated due to the interference between the tree-level and one-loop level
diagrams when there are complex Yukawa couplings and more than one
RHN. In the above expression of the {\it CP} asymmetry
parameter, the first term represents interference between tree-level
and vertex correction diagrams, while in the second term, a self-energy correction diagram is involved. The resonant enhancement occurs
when the mass splitting between the two sterile states 
$M_i - M_j \simeq \frac{\Gamma_j}{2}$. In this case, the prefactor in the self-energy correction term becomes proportional to $\Gamma_j^{-1}$.
\begin{figure*}[t]
\centering
\includegraphics[width = 0.45\textwidth]{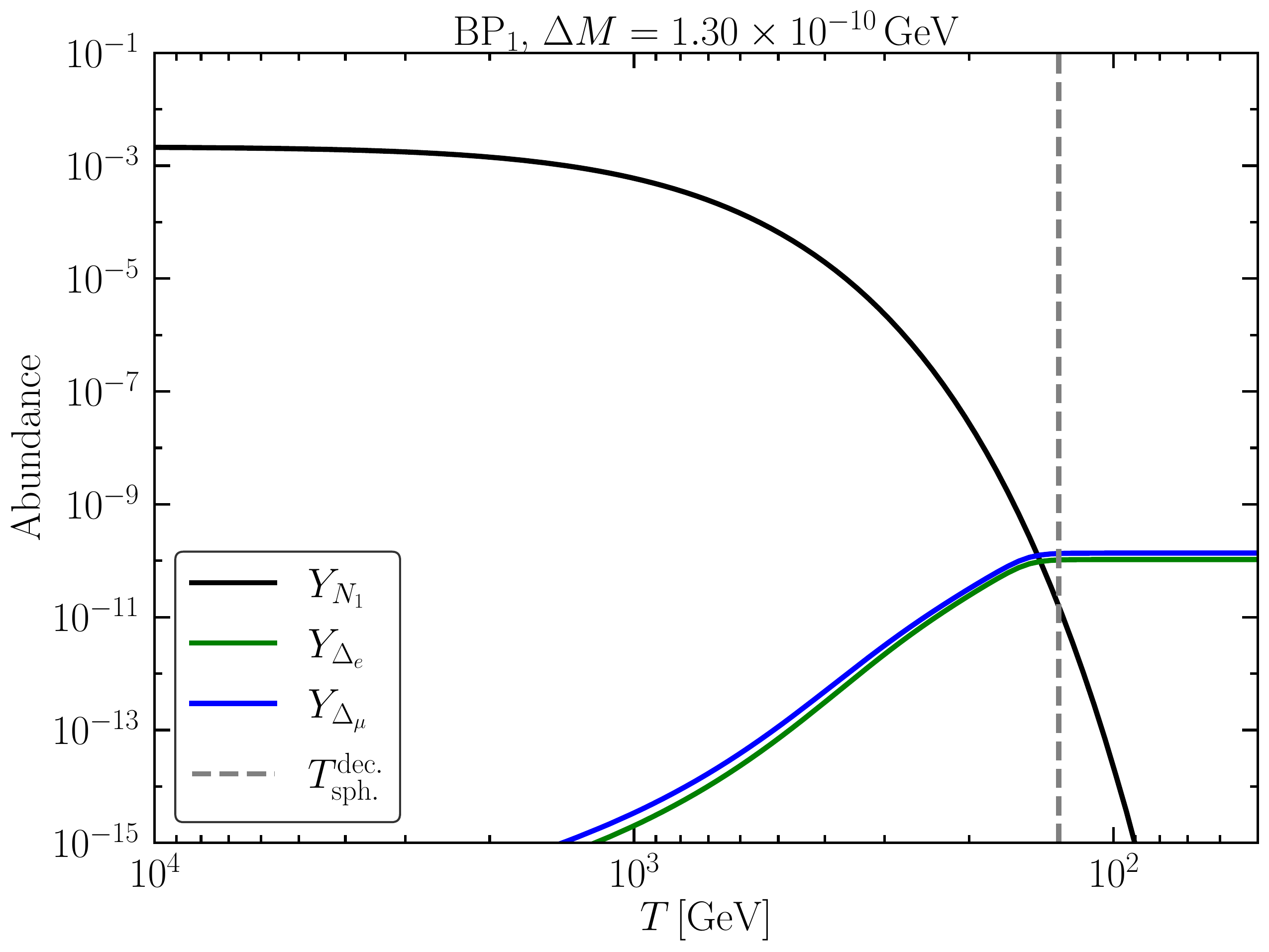}
\includegraphics[width = 0.45\textwidth]{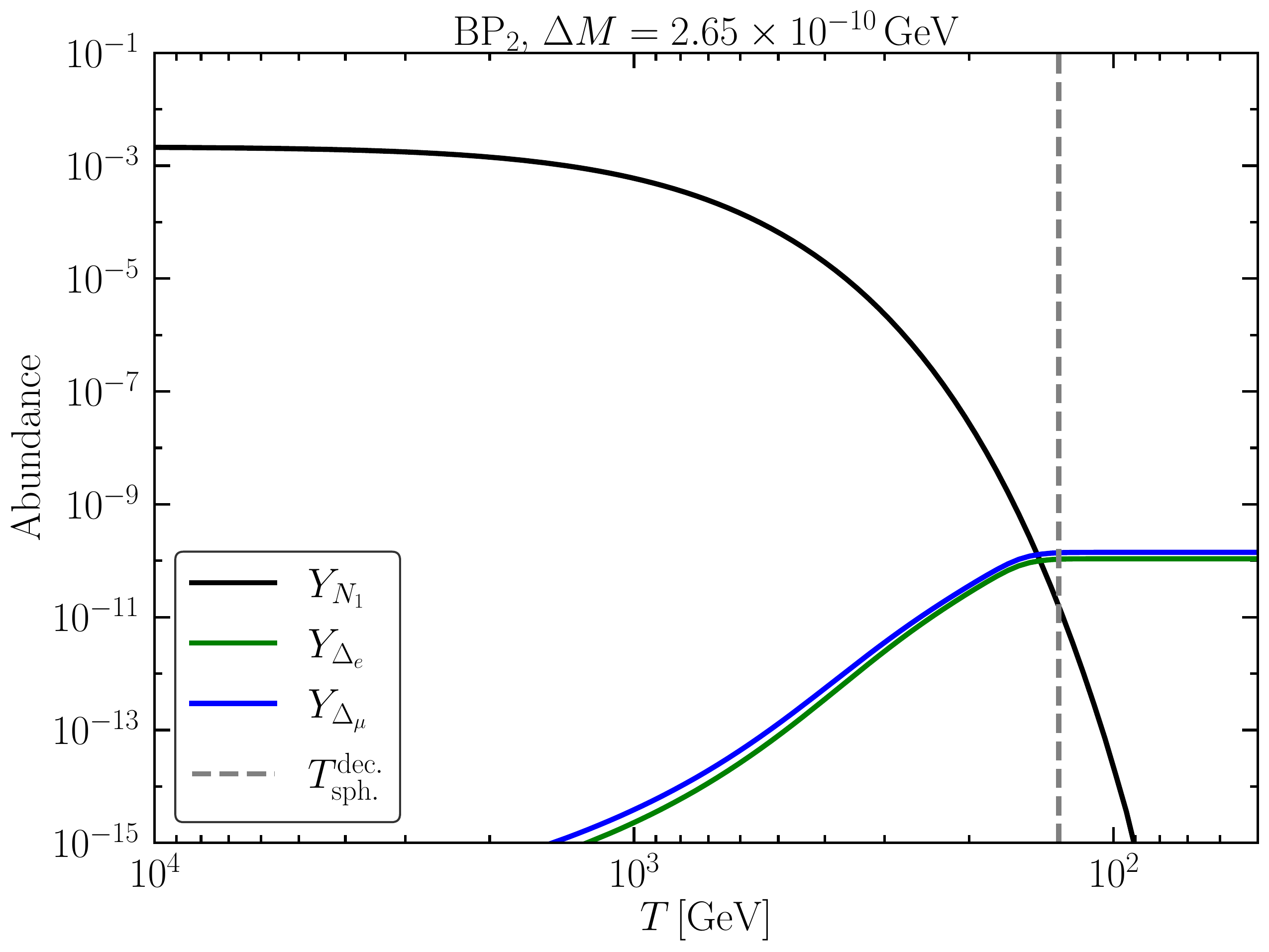}
\includegraphics[width = 0.45\textwidth]{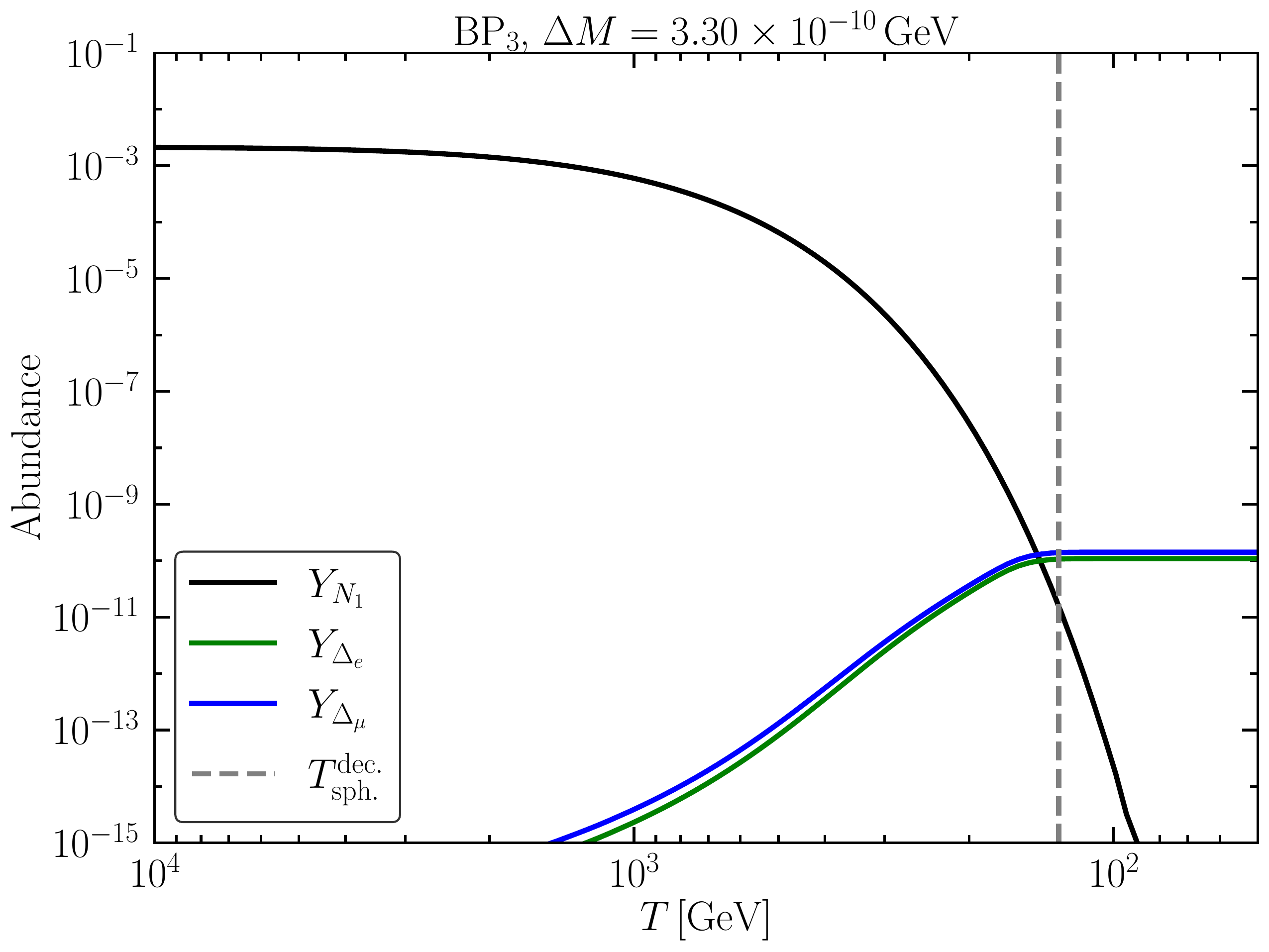}
\includegraphics[width = 0.45\textwidth]{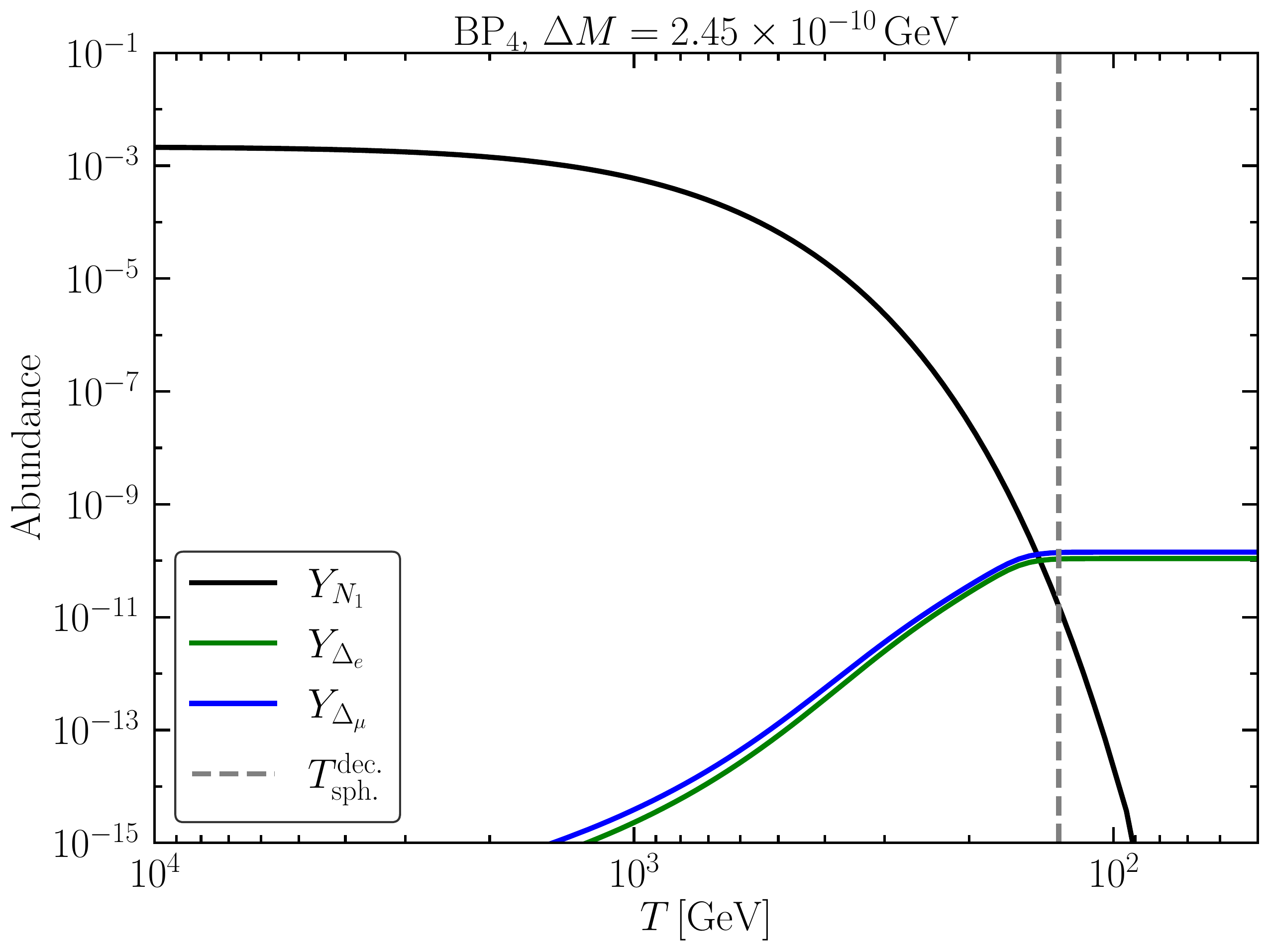}
\includegraphics[width = 0.45\textwidth]{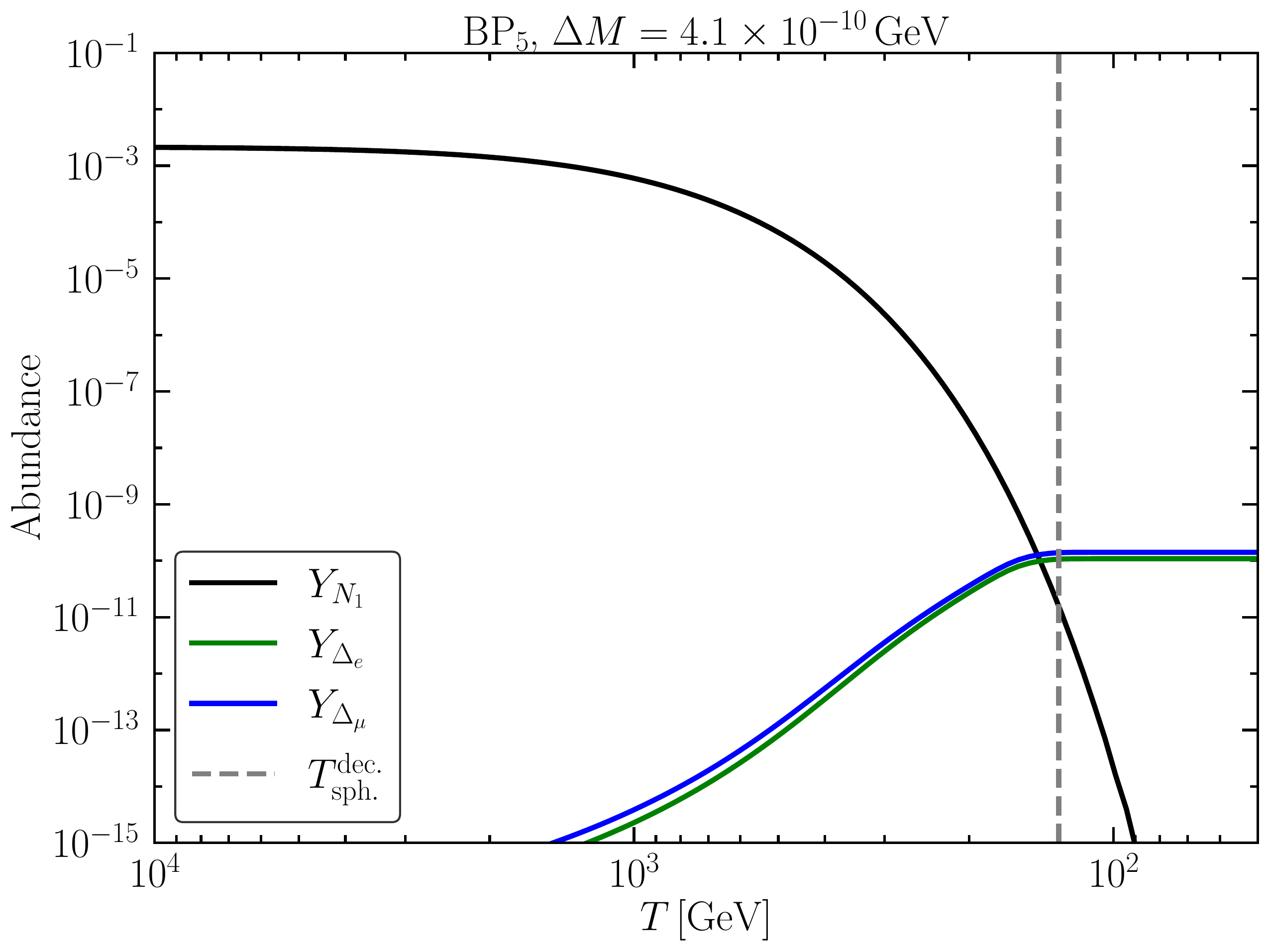}
\includegraphics[width = 0.45\textwidth]{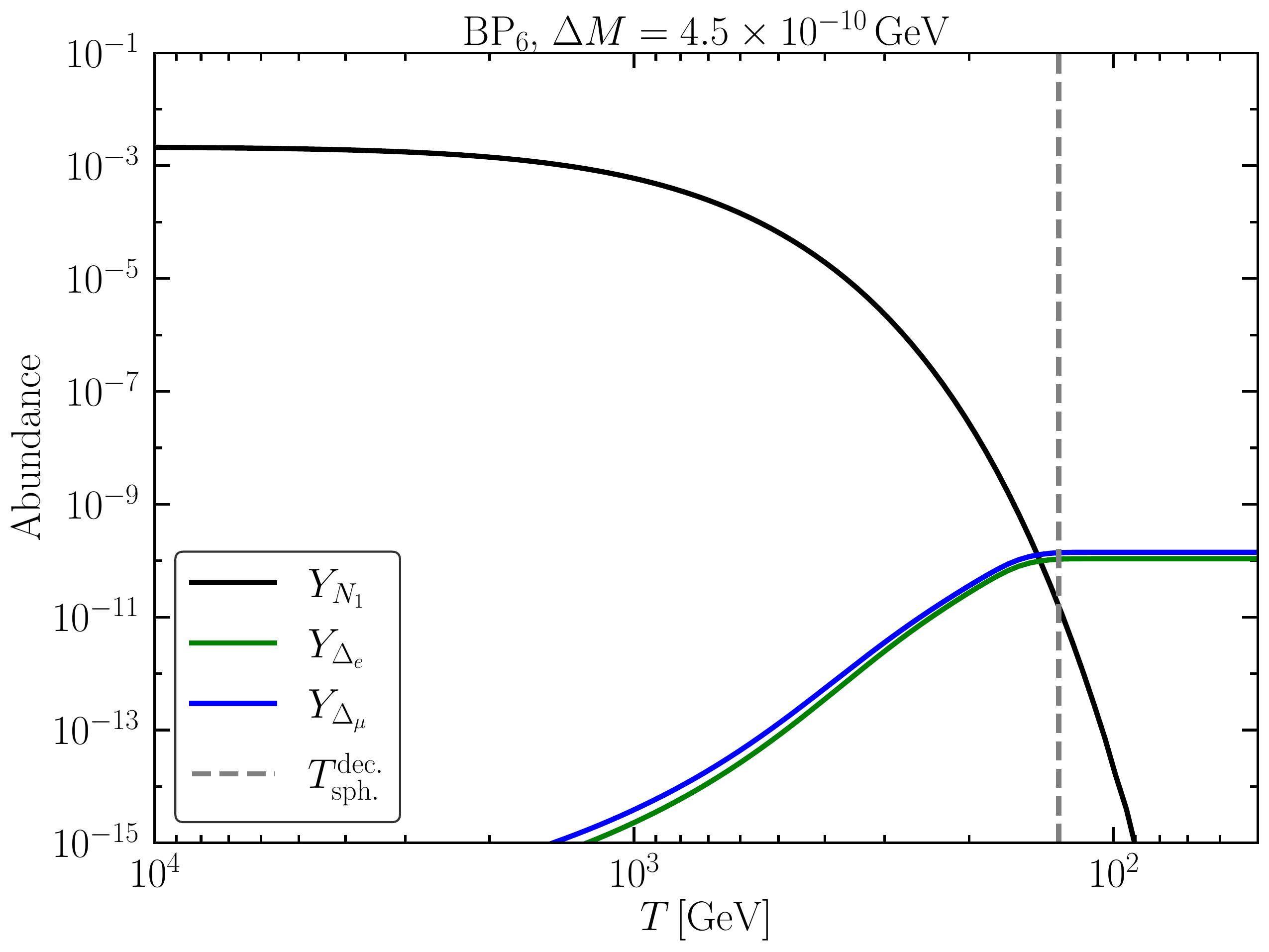}
\caption{Evolution of comoving number densities with $T$ for the six benchmark points. 
The gray dashed vertical line in each panel denotes the sphaleron
decoupling temperature $T_{\rm sph.}^{\rm dec.} = 130\,\rm GeV$.
}
\label{fig:line_plot}
\end{figure*}
In Fig.\,\ref{fig:CP_asymmetry}, the {\it CP} asymmetry parameter is demonstrated with the mass splitting $\Delta M \equiv M_2 - M_1$
for $M_1 = 3$ TeV. The solid lines represent the asymmetry generated
from the decay of $N_1$ into two different lepton flavors like
$e$, and $\mu$. The dashed lines indicate similar quantities
for the decay of the other RHN $N_2$. We would like to point out that there is a critical value of
$\Delta{M}_{ij} \equiv M_i - M_j$ below which the asymmetry parameter behaves
oppositely with the mass splitting. This happens when
$(1-x_{ji})^2 \ll x_{ji}\,\mathcal{Y}_{ji}$
(i.e., $M_i - M_j < \frac{\Gamma_j}{2}$), and in this case,
the prefactor becomes proportional to $\Delta{M}_{ij}$ i.e.,
$(1-x_{ji})/x_{ji}{\mathcal{Y}_{ji}} \sim \Delta{M}_{ij} M_i/\Gamma^2_j$.
This nature is seen in Fig.\,\ref{fig:CP_asymmetry}
where the critical value of the mass splitting is $\sim 10^{-10}$ GeV
for $M_1 = 3$ TeV.

The asymmetry parameter 
also depends on $\theta_R$ and
$\theta_I$ since $Y_{\rm D}$ depends on the $R$ matrix given in Eq.\,\eqref{eq:Rmatrix}. In order to
to understand the nontrivial dependence of $\epsilon_{i \alpha}$
on $\theta_R$ and $\theta_I$, in
Fig.\,\ref{fig:CP_asymmetry_vs_thetaI}, we have shown the variation of the asymmetry parameter
as a function of $\theta_I$ for $\theta_R = 1\,\rm rad$. In this figure, the other parameters are 
fixed according the benchmark point 1( BP1), as mentioned in Table\,\ref{tab:BP}. As one can see from the figure, the {\it CP} asymmetry is peaked when $\theta_I < 0$.
Thus, in the rest of our analysis, we have fixed $\theta_R = 1\,\rm rad$ and $\theta_I = -1\,\rm rad$. Let us note in passing that
we have also checked the dependence of $\epsilon_{i\alpha}$
on $\theta_R$, and find that the dependence is mild.

\begin{figure*}[t]
\centering
\includegraphics[width = 0.45\textwidth]{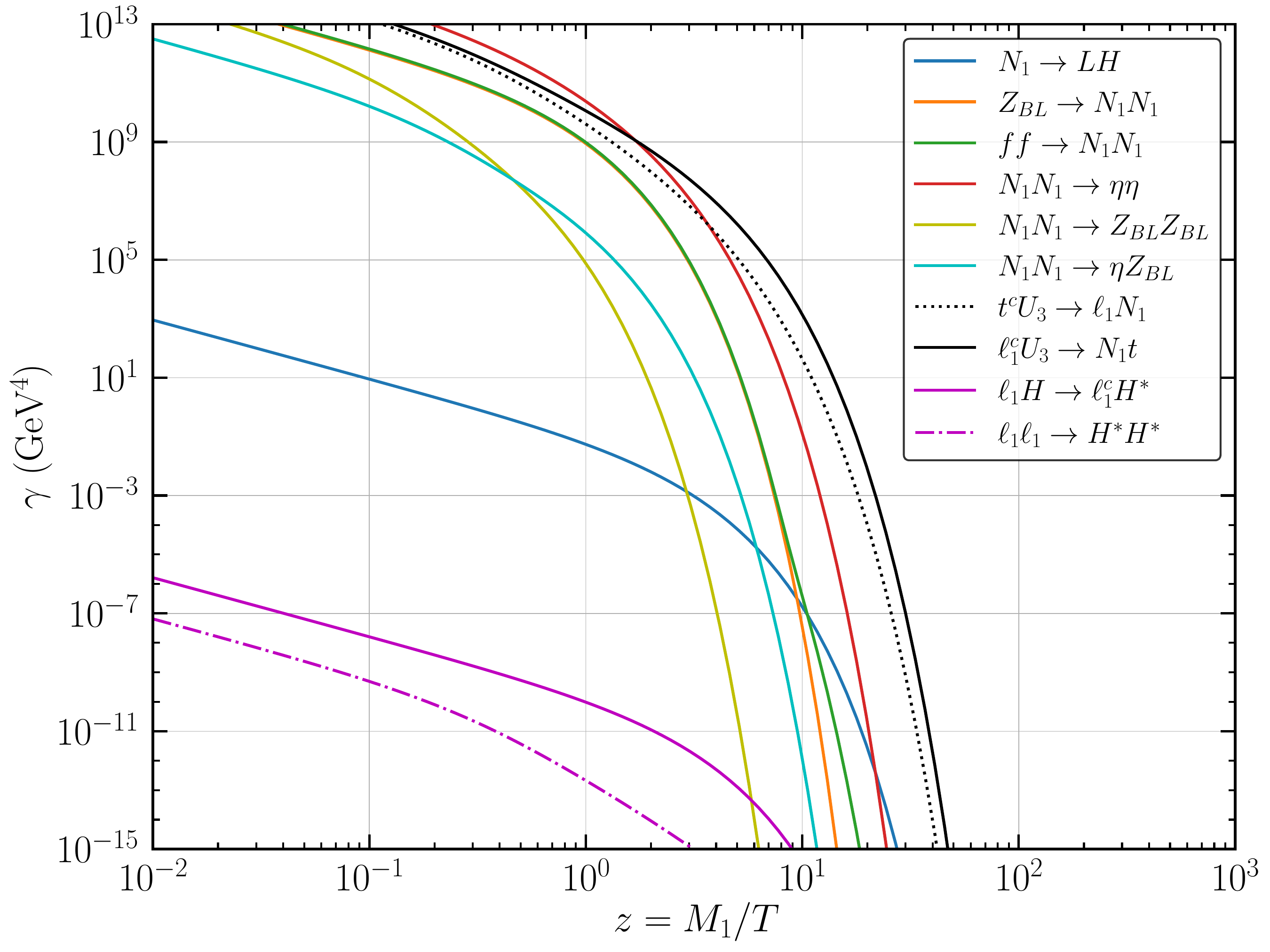}
\includegraphics[width = 0.45\textwidth]{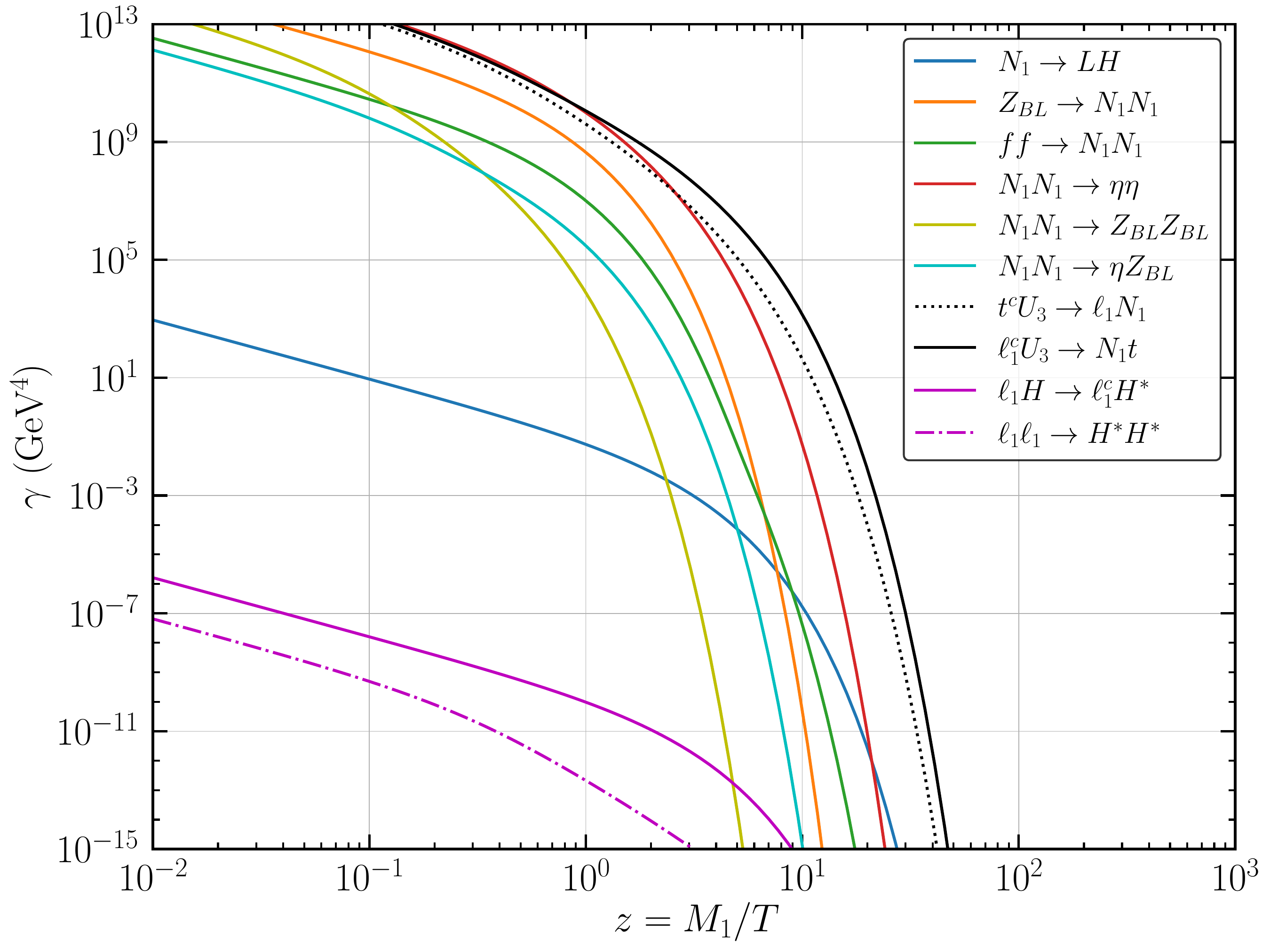}
\caption{$\gamma$ as a function of $z$ for the
processes relevant in the Boltzmann equation
for BP1 (left panel) and BP3 (right panel).}
\label{fig:gamma_plot}
\end{figure*}

Now, to calculate the baryon asymmetry, we have solved the Boltzmann equations for two RHNs and
three lepton flavors [Eqs. \eqref{eq:BE_N1} and \eqref{eq:lepton_assymetry_BE}]
numerically using Eqs.\,\eqref{eq:RHN_decay_width}\,--\,\eqref{eq:new_scattering_gamma} and \eqref{eq:asymmetry_para}
to obtain the respective comoving abundance. 
In Fig.\,\ref{fig:comparison_plot}, we have demonstrated\footnote{$Y_{\Delta_{\tau}}$
vanishes as the {\it CP} asymmetry parameters for the third-generation leptons are identically zero.} 
$Y_\Delta \equiv Y_{{\Delta}_e}+Y_{{\Delta}_\mu}$ for three different values
of $M_1=1$ TeV (blue), 2 TeV (green) and 3 TeV (red)
respectively. From this plot, it is clearly seen
that $Y_{\Delta}$ saturates before the sphaleron
decoupling temperature ($\simeq 130$ GeV) only when
$M_1 \gtrsim 3$ TeV. For $M_1 \lesssim 3$ TeV, the lepton
asymmetry saturates after the freeze-out of the sphaleron,
and therefore, the former will not be converted
into the baryon asymmetry efficiently. In the same
plot, for completeness, we have also illustrated
the abundance of $N_1$ for the three chosen
values of $M_1$. Therefore in the rest of our analysis, we fix
$M_1= 3\,\rm TeV$,
and $M_3 = 10\,\rm TeV$.

In Fig.\,\ref{fig:line_plot},
we have shown our results for six benchmark points listed in
Table \ref{tab:BP}. The black line represents the comoving
abundance for the lightest RHN $N_1$. Since we are considering
a resonant enhancement in the {\it CP} asymmetry parameter, $N_2$ is almost degenerate to $N_1$, and
hence its evolution is identical to the lightest RHN; therefore,
we have not shown $Y_{N_2}$ separately. 
The comoving abundances for
$Y_{\Delta_{\alpha}}$ ($\alpha = e$ and $\mu$) are
indicated by green and blue lines, respectively. 
As we have seen for the asymmetry parameter (Fig.\,\ref{fig:CP_asymmetry}), a similar nature is
reflected in the comoving abundance
also, i.e., $Y_{\Delta_{e}} \gtrsim Y_{\Delta_{\mu}}$.

To understand the relative magnitudes of the various decay, inverse
decay, and $2\rightarrow 2$ scatterings in the collision terms,
we have plotted $\gamma$ as a function of $z$ in Fig.\,\ref{fig:gamma_plot}.
Finally, the baryon-to-photon ratio ($\eta_{B}$), an experimentally
obtained physical quantity, can be determined from $Y_{\Delta_{\alpha}}$ using Eq.\,\eqref{eq:etaB}. For all
the benchmark points, $\eta_B$ is obtained within the $2\sigma$ range of experimental measurement by suitably adjusting the
mass gap between the two quasidegenerate RHNs. 
As an example, for BP1, BP2, and BP3, the mass gap $\Delta M = 1.3\times 10^{-10}$ GeV, $2.65\times 10^{-10}$ GeV and $3.3\times 10^{-10}$
GeV, respectively while
$\Delta M = 2.45\times 10^{-10}$ GeV,
$4.1\times 10^{-10}$ GeV and $4.5\times 10^{-10}$ GeV, respectively
for the other three benchmark points.

In the next section, we will discuss the production of the SGWB from an FOPT and 
the prospects of probing the parameter space of resonant flavored leptogenesis in future GW detectors.

\section{FIRST-ORDER PHASE TRANSITION AND GRAVITATIONAL WAVE}
\label{sec:FOPT_GW}
In SM, the phase transition is a crossover type, and thus presence 
of an extra scalar field in the scalar sector makes the dynamics more intriguing. In this section, 
we discuss the dynamics of the first-order transition when the scalar sector contains a complex
scalar field with nonzero $U(1)_{\bl}$ quantum number. The presence of an extra scalar field
and a $U(1)_{\bl}$ gauge boson
can generate GW signals, which can be detected in upcoming GW detectors. In the following sections,
we discuss our scalar potential, dynamics of first-order phase transition, and GW signals. First, 
we start our discussion with a description of the scalar potential. 
\subsection{Scalar potential in the thermal background}
\label{sec:scalar_potential_th_bkg}
The tree-level potential in terms of $\zeta_1$ and $\eta_1$ is as follows:
\bal
\label{eq:Vtree}
V_{\rm Tree} 
= -\dfrac{\mu_H^2}{2} \zeta_1^2 - \dfrac{\mu_\Phi^2}{2} \eta_1^2
+\dfrac{\lambda_H}{4} \zeta_1^4 + \dfrac{\lambda_\Phi}{4} \eta_1^4
-\dfrac{\lambda^\prime}{4}\zeta_1^2 \eta_1^2\,\,.
\eal
The one-loop corrected potential
of Eq.\,\eqref{eq:Vtree} in the thermal background has two 
parts, (i) one-loop corrected potential at zero temperature,
known as the Coleman-Weinberg (CW) potential ($V_{\rm CW}$) \cite{Coleman:1973jx} and (ii) 
the temperature-dependent part of the one-loop correction ($V_{T\ne 0}$). First, we will discuss
the CW potential for our model. The CW potential of our model in the Landau gauge is given by
\begin{widetext}
\bal
\label{eq:VCW}
V_{\rm CW}   &= \dfrac{1}{64\pi^2} 
        \left[
        \mbar_{\zeta}^4 \left( \ln \left(\dfrac{\mbar_{\zeta}^2}{\mu^2}\right) -\dfrac{3}{2} + C_{\rm UV}\right)
        +\mbar_{\eta}^4 \left( \ln \left(\dfrac{\mbar_{\eta}^2}{\mu^2}\right) -\dfrac{3}{2} +  C_{\rm UV}\right) 
        +2\mbar_{G^+}^4 \left( \ln \left(\dfrac{\mbar_{G^+}^2}{\mu^2}\right) -\dfrac{3}{2} +  C_{\rm UV}\right)\right. \nn\\
        &\left.+\mbar_{\zeta_2}^4 \left( \ln \left(\dfrac{\mbar_{\zeta_2}^2}{\mu^2}\right) -\dfrac{3}{2} + C_{\rm UV}\right)
        +\mbar_{\eta_2}^4 \left( \ln \left(\dfrac{\mbar_{\eta_2}^2}{\mu^2}\right) -\dfrac{3}{2} + C_{\rm UV}\right)
        +6\,\mbar_W^4 \left( \ln \left(\dfrac{\mbar_W^2}{\mu^2}\right) -\dfrac{5}{6} + C_{\rm UV}\right) \right.\nn \\
        &\left. +3\,\mbar_Z^4 \left( \ln \left(\dfrac{\mbar_Z^2}{\mu^2}\right) -\dfrac{5}{6} + C_{\rm UV}\right) 
        +3\,\mbar_{\zbl}^4 \left( \ln \left(\dfrac{\mbar_{\zbl}^2}{\mu^2}\right) -\dfrac{5}{6} + C_{\rm UV}\right)
        -12\,\mbar_t^4 \left( \ln \left(\dfrac{\mbar_t^2}{\mu^2}\right) -\dfrac{3}{2} + C_{\rm UV}\right) \right.\nn \\
        &\left.-2\sum_{i=1}^3\,
        \mbar_{i}^4 \left( \ln \left(\dfrac{\mbar_{i}^2}{\mu^2}\right) -\dfrac{3}{2} + C_{\rm UV}\right)
        \right] \,\,,
\eal
\end{widetext}
where $\mu$ is the renormalization scale and we consider $\mu = \vbl$. The field-dependent masses of the SM and BSM degrees of 
freedoms are denoted by $\mbar$, and they are listed in Appendix \ref{AppD:masses}. We would like to mention that the form of Eq.\,\eqref{eq:VCW} is gauge dependent,
and gauge-independent study is beyond the scope of this work (See \cite{Fukuda:1975di, Nielsen:1975fs} for a gauge-independent analysis). We confined our analysis in the Landau gauge $(\xi = 0)$ 
so that the contributions of the ghosts were absent in $V_{\rm CW}$
\cite{Delaunay:2007wb}. $C_{\rm UV} = -(2/\epsilon - \gamma + \log 4\pi$) in the CW potential contains the 
divergent part, which will be removed by introducing the following counterterm Lagrangian:
\bal
 \label{eq:CT_pot}
            V_{\rm CT} = -\dfrac{\delta_{\mu_H}}{2} \zeta_1^2 - \dfrac{\delta_{\mu_\Phi}}{2} \eta_1^2 + \dfrac{\delta_{\lambda_H}}{4} \zeta_1^4
            +\dfrac{\delta_{\lambda_\Phi}}{4} \eta_1^4 - \dfrac{\delta_{\lambda^\prime}}{4} \zeta_1^2 \eta_1^2\,\,.
\eal
We impose the following renormalization conditions to determine the unknown coefficients of Eq.\,\eqref{eq:CT_pot}.
    \bal
    \label{eq:CT_cond}
        &\dfrac{\partial(V_{\rm CW} + V_{\rm CT})}{\partial \zeta_1} = 0,\,~~
        \dfrac{\partial(V_{\rm CW} + V_{\rm CT})}{\partial \eta_1}  = 0 \,,\nn\\
        &\dfrac{\partial^2(V_{\rm CW} + V_{\rm CT})}{\partial \zeta_1^2}  = 0,~~
        \dfrac{\partial^2(V_{\rm CW} + V_{\rm CT})}{\partial \eta_1^2}  = 0\,,\nn\\
        &\dfrac{\partial^2(V_{\rm CW} + V_{\rm CT})}{\partial \zeta_1 \partial \eta_1}  = 0\,\,,
\eal
\allowdisplaybreaks
where the first two conditions ensure that the loop correction does not alter the tree-level vacua and the last 
three conditions make the tree-level scalar mass matrix unaltered. Solving Eq.\,\eqref{eq:CT_cond}, we obtain
\bal
        & \delta_{\mu _H} = \dfrac{3 V_{\rm CW}^{\zeta_1} - \vbl V_{\rm CW}^{\zeta_1 \eta_1} - \vh V_{\rm CW}^{\zeta_1 \zeta_1}}{2 \vh}\,,\nn\\
         &\delta_{\mu _\Phi} = \dfrac{3 V_{\rm CW}^{\eta_1} - \vh V_{\rm CW}^{\zeta_1 \eta_1} - \vbl V_{\rm CW}^{\eta_1 \eta_1}}{2 \vbl}\,,\nn\\
        & \delta_{\lambda_H} = \dfrac{V_{\rm CW}^{\zeta_1} - \vh V_{\rm CW}^{\zeta_1 \zeta_1}}{2 \vh^3}\,,\nn\\
        &\delta_{\lambda_\Phi} = \dfrac{V_{\rm CW}^{\eta_1} - \vbl V_{\rm CW}^{\eta_1 \eta_1}}{2 \vbl^3}\,,~~
        \delta_{\lambda^\prime} = \dfrac{V_{\rm CW}^{\zeta_1 \eta_1}}{\vh \vbl}\,\,,
    \eal
where $V^{x, y}_{\rm CW} \equiv \dfrac{\partial^2 V_{\rm CW}}{\partial x \partial y}$ is calculated at the vacuum configuration. Note that the Goldstone
bosons are massless at the physical vacuum, and it makes the calculation unstable once we calculate the derivatives of $V_{\rm CW}$ at the vacuum configuration.
In our calculation, this problem is circumvented by adding a regulator mass $m_{\rm IR}$ to the field-dependent Goldstone boson mass $\mbar_G$ as $\mbar^2_{G} \to \mbar_G^2 + m_{\rm IR}^2$. Here,
we took $m_{\rm IR}^2 = 10\,\rm GeV^2$, which is much smaller than the mass scale of the other degrees of freedom. However, in our numerical analysis, 
the derivatives of $V_{\rm CW}$ are calculated using the finite-difference method \cite{Press:1992zz}, and the choice of the increment of the independent variable serves as a regulator
and it prevents the function $V_{\rm CW}$ from encountering the singularity at the physical vacuum.

In the thermal background, the scalar potential can be derived from the following equation \cite{Quiros:1999jp}:
\bal
\label{eq:VT_master}
V^{\rm total}_{1 -{\rm loop}} = \sum_B\dfrac{g_B T}{2} \sumint_{n_b=0}^\infty \dfrac{d^3 \vec{k}}{(2 \pi)^3} \ln (\omega_{n_b}^2 + \omega^2) \nn\\
- \sum_F\dfrac{g_F T}{2} \sumint_{n_f=0}^\infty \dfrac{d^3 \vec{k}}{(2 \pi)^3} \ln (\omega_{n_f}^2 + \omega^2) \,\,,
\eal
where the summation over $B$ and $F$ runs over all bosons and fermions. $T$ is the plasma temperature and  $\omega^2 = |\vec{k}|^2 + \mbar^2$. $\omega_b = 2 \pi n_b T$ and
 $\omega_f = (2n_f + 1) \pi T$ are Matsubara frequencies for the bosons, fermions, respectively. The coefficients $g_B = 1, 2, 3, 4, 2$
for the real scalar, complex scalar, massive gauge boson, Dirac fermion, and Majorana fermion, respectively. 
After performing the Matsubara frequency sum, one can write Eq.\,\eqref{eq:VT_master} as
\bal
V^{\rm total}_{1 -{\rm loop}}(\zeta_1,\,\eta_1) = V_{\rm CW} (\zeta_1,\,\eta_1) + V_{T \ne 0}(\zeta_1,\,\eta_1)\,\,,
\eal
where
\bal
V_{T \ne 0}(\zeta_1,\,\eta_1) = \dfrac{T^4}{2 \pi^2}
\left[
\sum_B g_B J_B \left(\dfrac{\mbar_B^2}{T^2}\right) - 
\sum_F g_B J_F \left(\dfrac{\mbar_F^2}{T^2}\right)
\right]\,\,.
\eal
The explicit form of $J_B(x_B)$ and $J_F(x_F)$ is as follows:
\bal
&J_B(x) = \int_0^\infty dz \,z^2\,\ln\left(1 - \exp(-\sqrt{z^2 + x})\right)\,\,\nn\\
&J_F(x) = \int_0^\infty dz\,z^2\, \ln\left(1 + \exp(-\sqrt{z^2 + x})\right)\,\,.
\eal
Lastly, the bosonic sum
in Eq.\,\eqref{eq:VT_master} becomes infrared (IR) divergent due to the presence of the $n_{b}=0$ contribution in the first term
and the perturbative expansion becomes unreliable. In order to circumvent this IR divergence, we need to resum the thermal mass of the bosonic 
degrees of freedom. We follow
the Arnold-Espinosa method \cite{Arnold:1992rz} and
add the following to the scalar potential:\footnote{Resummation of the thermal mass of fermionic fields is not 
required since the fermionic summation in Eq.\,\eqref{eq:VT_master}
is not IR divergent because $\omega_{n_f} \neq 0$ for $n_f = 0$.}. 
\bal
\label{eq:V_daisy}
V_{\rm daisy} = \sum_B \dfrac{g_B T}{12 \pi} \left[\mbar_B^3 - \mbar_{B,T}^3\right]\,\,,
\eal
where the summation runs over all the bosonic degrees of freedom, and $\mbar_{B,T}$ is the field-dependent mass of the bosonic fields in the thermal background. The explicit forms
of $\mbar_{B,T}$ for all the bosonic degrees of freedom are given in Appendix \ref{AppD:masses}. Let us note that, for  gauge bosons, the value of $g_B$ in Eq.\,~\eqref{eq:V_daisy} is 1 since only longitudinal modes acquire thermal mass.

Thus, the total scalar potential in our case is given by
\bal
\label{eq:total_potential}
V_{\rm total} = V_{\rm tree} + V_{\rm CW} + V_{\rm CT} + V_{T \ne 0} + V_{\rm daisy}\,\,.
\eal
In the next section, we will discuss some important features of Eq.\,\eqref{eq:total_potential} and the production of the GW signal.
\subsection{FOPT dynamics and GW signal}
\label{sec:FOTP_and_GW}
Before delving into our numerical results, we use our tree-level and thermal potential in order to understand the FOPT dynamics
in the presence of an extra scalar with a nonzero $U(1)_{\bl}$ charge. Neglecting $V_{\rm CW}$ and $V_{\rm daisy}$, in the
high-temperature limit, we can write Eq.\,\eqref{eq:total_potential} as
\bal
\label{eq:Vtotal_high_T}
V_{\rm total} &\simeq \dfrac{1}{2} (-\mu_H^2 + c_H T^2) \zeta_1^2
+\dfrac{1}{2} (-\mu_\Phi^2 + c_\Phi T^2) \eta_1^2 \nn\\
&-E_H \zeta_1^3 T - E_{\Phi} \eta_1^3 T
+ \dfrac{\lambda_H}{4} \zeta_1^4 + 
\dfrac{\lambda_\Phi}{4} \eta_1^4 -
\dfrac{\lambda^\prime}{4} \zeta_1^2 \eta_1^2\,,
\eal
where 
\bal
&c_H = \left(\dfrac{g_1^2}{16} + \dfrac{3 g_2^2}{16} + \dfrac{y_t^2}{4} + \dfrac{\lambda_H}{2} - \dfrac{\lambda^\prime}{12}\right)\,\,,\nn\\
&c_\Phi = \left(\gbl^2 + \dfrac{y_N^2}{8} + \dfrac{\lambda_\Phi}{3} - \dfrac{\lambda^\prime}{6}\right)\,\,,\nn\\
&E_H = \dfrac{m_W^3}{2 \pi \vh^3} + \dfrac{m_Z^3}{4\pi \vh^3}\,\,,~~
E_\Phi = \dfrac{m_{\zbl}^3}{4 \pi \vbl^3}\,\,.
\eal

At high temperature, both EW and $U(1)_{\bl}$ symmetries are restored
due to the temperature-dependent coefficients of $\zeta_1^2$ and
$\eta_1^2$. However, as the Universe cools down, a phase transition
occurs. First, we start our discussion with $\gbl \to 0$ which
implies that $E_\Phi$ vanishes. The absence of $E_\Phi$ implies
that the FOPT occurs in the $\zeta_1$ direction only. In this
limit, a first-order electroweak phase transition (FOEWPT) occurs when
\bal
\dfrac{v_c}{T_c} = \dfrac{8 E_{H} \lambda_\phi}{4 \lambda_H \lambda_\Phi - {\lambda^\prime}^2}\,\,,
\label{eq:vcbytc}
\eal
where $v_c$ is the EW VEV at critical temperature $T_c$. 
If we substitute all the quartic couplings given in Eq.\,\eqref{eq:vcbytc}, we will find that the ratio $v_c/T_c$ is 
greater than 1 when the mixing angle is large and $m_{\eta} < m_{\zeta}$. However, in our parameter space of interest, 
the mass of the extra scalar is larger than the SM Higgs and
as a result, FOEWPT is not possible in our scenario. 
In spite of this, we can still have an FOPT in the $\eta_1$ direction,
thanks to $\gbl \ne 0$, and as a result, SGWB can be generated. 
In the next section, we will discuss the SGWB signal originated from an FOPT and present our numerical results.
\subsection{Stochastic gravitational wave background from the FOPT}
\label{sec:SGWB}
The amplitude of the SGWB emitted from the FOPT has three sources, and the total amplitude is
\bal
\label{eq:GW_amplitude}
\Omega_{\rm GW} h^2 = \Omega_{\rm coll} h^2 + \Omega_{\rm sw} h^2 + \Omega_{\rm MHD} h^2\,\,,
\eal
where individual contributions from the bubble-wall collision ($\Omega_{\rm coll} h^2$),
sound waves in the plasma using an envelope approximation ($\Omega_{\rm sw} h^2$), and 
magnetohydrodynamic turbulence in the plasma ($\Omega_{\rm turb} h^2$) are as follows \cite{Caprini:2015zlo}.

The contribution to the GW amplitude from the bubble-wall collision is
\bal
\label{eq:bubble_amplitude}
\Omega_{\rm GW} h^2 
&= 1.67 \times 10^{-5} \left(\dfrac{H_*}{\beta}\right)^2 \left(\dfrac{\kappa \alpha}{1 + \alpha}\right)^2
\left(\dfrac{100}{g_*(T_n)}\right)^{1/3} \nn\\
&\times \left(\dfrac{0.11 v_w^3}{0.42 + v_w^2}\right) 
\left(\dfrac{3.8 (f/f_{\rm col})^{2.8}}{1 + 2.8(f/f_{\rm col})^{3.8}}\right)\,\,,
\eal
with the peak frequency of the spectrum 
\bal
\label{eq:bubble_peak}
f_{\rm col} &= 1.65 \times 10^{-5}\,{\rm Hz} \left(\dfrac{0.62}{1.8 - 0.1v_w + v_w^2}\right)
\left(\dfrac{\beta}{H_*}\right) \left(\dfrac{T_n}{100\,\rm GeV}\right) \nn\\
&\times \left(\dfrac{g_*(T_n)}{100}\right)^{1/6}\,\,.
\eal
The contribution of the sound wave after the bubble collision is
\bal
\label{eq:sw_amplitude}
\Omega_{\rm sw} h^2 &= 2.65 \times 10^{-6} \left(\dfrac{H_*}{\beta}\right) \left(\dfrac{\kappa_v \alpha}{1 + \alpha}\right)^2
\left(\dfrac{100}{g_*(T_n)}\right)^{1/3} \nn\\
&\times v_w \left(\dfrac{f}{f_{\rm sw}}\right)^3 \left(\dfrac{7}{4 + 3 (f/f_{\rm sw})^2}\right)^{7/2}\,\,,
\eal
and in this case the peak frequency is
\bal
\label{eq:sw_peak}
f_{\rm sw} = 1.9 \times 10^{-5}\,{\rm Hz} \,\,v_w^{-1} 
\left(\dfrac{\beta}{H_*}\right) \left(\dfrac{T_n}{100\,\rm GeV}\right) \left(\dfrac{g_*(T_n)}{100}\right)^{1/6}\,\,.
\eal
Lastly, the contribution from $\Omega_{\rm MHD} h^2$ is given by
\bal
\label{eq:MHD_amplitude}
\Omega_{\rm MHD} h^2 &= 3.35 \times 10^{-4} \left(\dfrac{H_*}{\beta}\right) \left(\dfrac{\kappa_{\rm turb} \alpha}{1 + \alpha}\right)^{3/2}
\left(\dfrac{100}{g_*(T_n)}\right)^{1/3} \nn\\
&\left(\dfrac{v_w (f/f_{\rm turb})^3}{ \left[1 + (f/f_{\rm turb})\right]^{11/3}\left(1 + 8 \pi f/h_*\right)}\right)\,\,
\eal
where 
\bal
h_* = 1.65 \times 10^{-5}\,{\rm Hz} \left(\dfrac{T_n}{100\,\rm GeV}\right) \left(\dfrac{g_*(T_n)}{100}\right)^{1/6}\,\,,
\eal
and the peak frequency 
\bal
\label{eq:MHD_peak}
f_{\rm turb} = 2.7 \times 10^{-5}\,{\rm Hz}\,\,v_w^{-1}\left(\dfrac{\beta}{H_*}\right) \left(\dfrac{T_n}{100\,\rm GeV}\right) \left(\dfrac{g_*(T_n)}{100}\right)^{1/6}\,\,.
\eal
\begin{table*}[t]
    \centering
    \begin{tabular}{|c|c|c|c|c|c|c|c|c|}
    \hline
    & $m_{\zbl}$ & $\gbl$ & $m_{\eta}$ & $\sin \theta$ & $M_1$ & $\alpha$ & $\beta/H_*$ & $T_n$\\
   \hline
    ${\rm BP}_1$ & $11.40 $ & $1 .01\times 10^{-1}$ & $0.5$ & $0.1$ & $3.0$ & $1.16\times 10^{-1}$ & $13.60 \times 10^3$ & 1.36\\
    \hline
    ${\rm BP}_2$ & $13.00$ & $8.11 \times 10^{-2}$ & $0.5$ & $0.1$ & $3.0$ & $1.01 \times 10^{-1}$ & $19.81 \times 10^3$ & 1.98\\
    \hline
    ${\rm BP}_3$ & $13.36$ & $7.62\times 10^{-2}$ & $0.5$ & $0.1$ & $3.0$ & $8.97 \times 10^{-2}$ & $21.85\times 10^3$ & 2.18\\
    \hline
    ${\rm BP}_4$ & $11.10$ & $7.11 \times 10^{-2}$ & $0.5$ & $0.1$ & $3.0$ & $3.97 \times 10^{-2}$ & $28.51 \times 10^3$ & 2.85\\
    \hline
    ${\rm BP}_5$ & $11.90$ & $6.31\times 10^{-2}$ & $0.5$ & $0.1$ & $3.0$ & $3.51 \times 10^{-2}$ & $35.08 \times 10^3$ & 3.51\\
    \hline
    ${\rm BP}_6$ & $11.54$ & $5.96 \times 10^{-2}$ & $0.5$ & $0.1$ & $3.0$ & $2.82 \times 10^{-2}$ & $42.15\times 10^3$ & 4.21\\
    \hline
    \end{tabular}
    \caption{Benchmark points for the FOPT which satisfy all the collider constraints. All the dimensionful quantities are in units of TeV.}
    \label{tab:BP}
\end{table*}
The bubble-wall speed is given by\footnote{For 
$\alpha \gtrsim 0.1$, the wall speed becomes
ultrarelativistic \cite{Laurent:2022jrs,
Krajewski:2024gma, Gouttenoire:2021kjv}. However, it does not
change the GW prediction significantly, even if we
use the Chapman-Jouguet velocity 
given in Eq.\,\eqref{eq:v_wall}.}
\bal
\label{eq:v_wall}
v_w = \dfrac{\sqrt{1/3} + \sqrt{\alpha^2 + 2 \alpha/3}}{1+\alpha}\,\,,
\eal
and we use
\bal
\label{eq:kappa}
\kappa &= \dfrac{0.715 \alpha + (4/27)*\sqrt{3\alpha/2}}{1 + 0.715 \alpha},~\kappa_v = \dfrac{\alpha}{0.73 + 0.083\sqrt{\alpha} + \alpha},\nn\\
&\kappa_{\rm turb} = 0.1 \kappa_v\,\,.
\eal
\begin{figure}[hbt!]
\centering
\includegraphics[width = 0.45\textwidth]{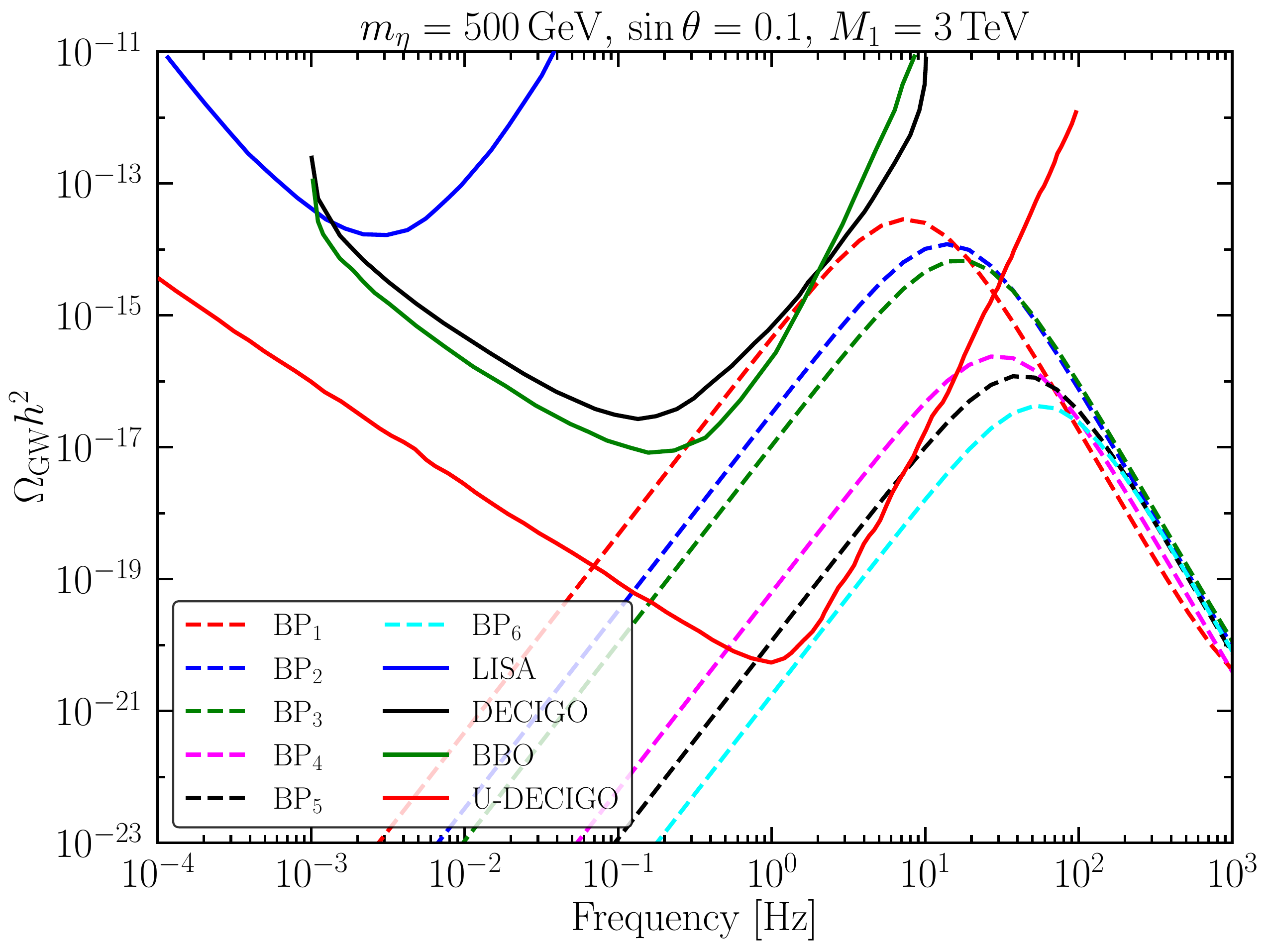}
\caption{Frequency spectrum of SGWB for six benchmark points tabulated in Table\,\ref{tab:BP}.
Here, we fix  $m_{\eta} = 500\,{\rm GeV}$, $\sin \theta =0.1$.
The sensitivity curves of LISA, DECIGO, BBO, and U-DECIGO are shown by blue, black, green, and red lines, respectively. The data for the sensitivity curves 
are taken from \cite{Schmitz:2020syl, Ringwald:2020vei}.
}
\label{fig:GW_spectrum}
\end{figure}
In Eqs.\,\eqref{eq:bubble_amplitude} --\eqref{eq:MHD_amplitude}, $T_n$ is the bubble nucleation temperature, $\beta^{-1}$ is the duration of the FOPT, and
$\alpha$ is the latent heat released during the transition, and they are defined
as \cite{Ellis:2018mja}
\bal
&\alpha = \left(\dfrac{\Delta V_{\rm total} - T\dfrac{\partial \Delta V_{\rm total}}{\partial T}}{\rho_{\rm rad}}\right)_{T = T_n}\,\,,\nn\\
&\beta = H_*\left(T\dfrac{d}{dT} 
\left(\dfrac{S_E}{T}\right)\right)_{T = T_n}\,\,,
\eal
where $V_{\rm total}$ is given in Eq.\,\eqref{eq:total_potential},
$\Delta V_{\rm total}$ is the potential difference between the false and true
vacuum, $H_*$ is the Hubble parameter at $T_n$, and $S_E$ is the three-dimensional Eucledian 
action \cite{Linde:1981zj}. These three quantities depend on the model parameters, and to identify 
the parameter space for the FOPT,  we implemented our model in the publicly available \texttt{CosmoTransitions} 
code \cite{Wainwright:2011kj}. We scan $m_{\zbl}$
between 1 and 20 TeV and $\gbl$ between
$10^{-2}$ and $1.0$. The other parameters like the BSM scalar mass, and scalar 
mixing angle are kept fixed at  $m_{\eta} = 500$ GeV, $M_1 = 3\,\rm TeV$, and $\sin{\theta} = 0.1$, respectively. 

\begin{figure}
    \centering    \includegraphics[width=0.48\textwidth]{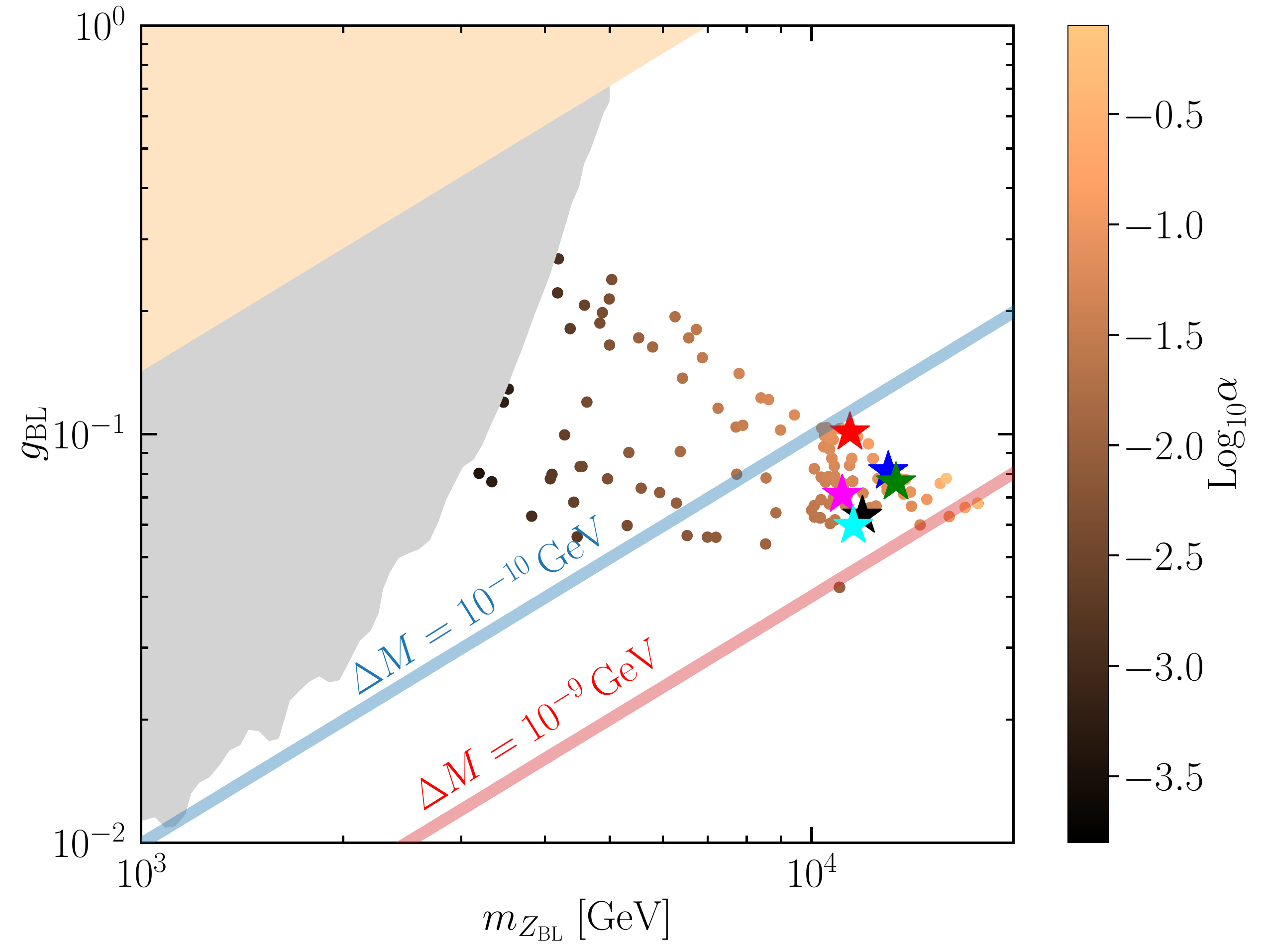}
    \caption{Parameter space in the $m_{\zbl} - \gbl$ plane for $m_{\eta} = 500\,{\rm GeV}$, $\sin \theta = 0.1$, $M_1 = 3\,\rm TeV$. 
    The scattered points represent the FOPT and the strength of the transition ($\alpha$) is depicted by the color bar. 
    The gray region is excluded from the dilepton search at LHC and the yellow region is disallowed from the measurement of 
    the $e^+e^- \to \ell^+ \ell^-$ search at the LEP. Observed baryon asymmetry at $2\sigma$ confidence interval 
    can be explained within the blue and red bands for $\Delta M = 10^{-10}\,\rm GeV$ and $\Delta M = 10^{-9}\,\rm GeV$ 
    respectively. The benchmark points given in Table\,\ref{tab:BP} are denoted by $\star$, and 
    their color code is the same as Fig.\,\ref{fig:GW_spectrum}.
    }
    \label{fig:constraint}
\end{figure}

In Fig.\,\ref{fig:GW_spectrum}, 
we show the frequency spectrum of the emitted SGWB for six benchmark points, which are tabulated in Table\,\ref{tab:BP}. 
These points are allowed from the collider constraints at 95\% CL., they can also explain the observed baryon asymmetry.
The sensitivity curves of LISA, DECIGO, BBO, and U-DECIGO are shown by the
blue, black, green, and red solid lines, respectively. As one can see from the figure that the amplitude of the emitted SGWB increases with $\alpha$. 
This is because a larger value of $\alpha$ 
implies that the amount of emitted latent heat is large, and thus the strength of the SGWB is large as well. However, 
the peak frequency anticorrelates with $\alpha$, and this can be understood as follows: For larger $\alpha$,
the energy difference between false and true vacuum is large, which implies that nucleation happens at a lower temperature. 
Since the peak frequency is directly proportional to $T_n$, it decreases with the increase in $\alpha$.

In Fig.\,\,\ref{fig:constraint}, we show the allowed region of parameter space in the $m_{\zbl} - \gbl$ plane 
where we fix $m_{\eta} = 500\,\rm GeV$, $\sin \theta = 0.1$,\,and $M_1  =  3\,\rm TeV$ and $M_3 = 10\,\rm TeV$, as discussed
earlier. The constraint from the dilepton search at the LHC is shown by the gray region and the corresponding 
LEP bound is shown by the light orange region as discussed in Sec.\,\ref{sec:constraints}. 
Here, one can see that for fixed $m_{\zbl}$, $\alpha$ is an increasing function
of $\gbl$. This is because for the FOPT in the $\eta_1$ direction,
the order parameter is proportional to $\gbl$ for the fixed value of
$m_{\zbl}$. As a result, we get larger $\alpha$ for higher values of $\gbl$. 
The blue and red bands denote the parameter space at which the observed baryon asymmetry can
be explained within the $2\sigma$ confidence interval. 
One can clearly see that
for\footnote{This mass hierarchy requires
$Y_{1,2} \simeq 0.08$, whereas $Y_3 \simeq 0.3$.}
$M_1 \simeq M_2=3\,\rm TeV$ and $M_3 = 10\,\rm TeV$,
$\zbl$ with mass $\sim 10\,\rm TeV$ and $\gbl \sim 0.1$ can 
produce an SGWB, and successfully explain the
observed baryon asymmetry of the Universe when the mass splitting
$\Delta M$ varies between $10^{-10}$ and $10^{-9}\,\rm GeV$. Importantly, despite being 
inaccessible to the current collider experiments, this region of parameter space can be tested
by the future GW detectors.
\section{SUMMARY AND CONCLUSIONS}
\label{sec:summary}
In this work, we have considered a gauged $\ubl$ extension of the SM which contains
three RHNs ($\ni$), one gauge boson $\zbl$, and a complex scalar $\Phi$
in the particle spectrum. One RHN per lepton flavor is the minimal requirement for 
gauge anomaly cancellation, while a massive $\zbl$ and a {\it CP}-even scalar
are natural consequences of spontaneously broken $\ubl$ gauge symmetry.
The presence of RHNs naturally accommodates sub-eV scale Majorana masses of SM neutrinos as well as
explains the observed baryon asymmetry of the Universe. Here, we consider the
TeV scale RHNs, and this can be achieved
by choosing the breaking scale of $\ubl$ and $\yf$ appropriately.

In this framework, {\it CP} asymmetry in the leptonic sector due to the decay of TeV scale 
RHNs with a hierarchical mass spectrum is not sufficient to produce correct baryon asymmetry. 
Thus, our scenario requires a significant boost in the {\it CP} asymmetry parameter which is eventually 
possible by virtue of two quasidegenerate RHNs through resonant enhancement. Moreover, the dynamics 
of RHNs is heavily influenced by the new gauge boson $\zbl$ and the scalar $\Phi$. We have 
calculated the abundance of RHNs taking into account all possible interaction channels involving 
$\zbl$, $\Phi$, and a top quark. Furthermore, in TeV scale, since all the Yukawa interactions 
of the SM leptons are in equilibrium, one has to consider the effect of each lepton flavor 
in the asymmetry calculation separately, requiring the solution of five coupled Boltzmann equations simultaneously. We
find the correct baryon asymmetry for a wide range of $m_{\zbl}$
which is beyond the reach of the present sensitivity of the LHC. 
  
Future GW detectors have the potential to probe a part of the allowed region of the parameter space 
which is inaccessible to the current colliders. The presence of the $\ubl$ breaking scalar
opens up the possibility to test the model under consideration by producing an SGWB via FOPT.
The symmetry-breaking scale is related to the frequency of emitted SGWB as well as the masses of an RHN and $\zbl$,
which plays a crucial role in explaining the baryon asymmetry of our Universe.
Leveraging the interconnection, we have investigated resonant leptogenesis in conjunction with the production of an SGWB from an FOPT. 
We identify the model parameter space that can explain the observed baryon asymmetry while producing an observable SGWB
that can be detected in BBO, DECIGO, and U-DECIGO. We also discuss current constraints from the collider and the 
complementarity between the collider searches and GW detection.
\section*{ACKNOWLEDGMENTS}
The work of S.G. was supported by the IBS under Project No. IBS-R018-D1.
We would like to thank Abhijit Kumar Saha, Benoit Laurent,
Carroll Wainwright, Francesco Costa, and Pankaj Borah for useful communication related to the \texttt{CosmoTransitions} package.

\newpage
\begin{widetext}
\appendix
\section{FORMULATION OF THE ${\cal A}$ MATRIX}
\label{AppA:A_matrix}
In this appendix, we will derive the structure of the $\cal A$ matrix used in Eq.\,\eqref{eq:A_matrix}. 

The asymmetry between the number density of particle $X$ and its antiparticle $\bar{X}$ can be written as
\bal
n_X - n_{\bar{X}} \simeq
\dfrac{T^3}{6}
\begin{cases}
\dfrac{2\,g_X\mu_X}{T}\, \text{for bosons} \\
\\
\dfrac{g_X\mu_X}{T}\, \text{for fermions}\,\,.
\end{cases}
\label{eq:asymmetry_th_eq}
\eal
Here, $g_X = \{6, \,3, \,3, \,2, \,1, \,2\}$ is the degree of the SM field content $\{Q_{L_i}, u_{R_i}, d_{R_i}, \ell_{L_i}, \ell_{R_i}, H\}$.

At the temperature when all the SM Yukawa couplings are in thermal equilibrium, it gives
\bal
&\mu_{\ell_{L_1}} - \mu_{\ell_{R_1}} - \mu_H = 0 \,,~~~~
\mu_{\ell_{L_2}} - \mu_{\ell_{R_2}} - \mu_H = 0\,,~~~~
\mu_{\ell{L_3}} - \mu_{\ell_{R_3}} - \mu_H = 0 \label{eq:lepton_yukawa}\,,\\
&\mu_{Q_{L_1}} - \mu_{d_{R_1}} - \mu_H = 0 \,,~~~~
\mu_{Q_{L_2}} - \mu_{d_{R_2}} - \mu_H = 0\,,~~~~
\mu_{Q_{L_3}} - \mu_{d_{R_3}} - \mu_H = 0 \label{eq:down_yukawa}\,,\\
&\mu_{Q_{L_1}} - \mu_{u_{R_1}} + \mu_H = 0\,,~~~~
\mu_{Q_{L_2}} - \mu_{u_{R_2}} + \mu_H = 0\,,~~~~
\mu_{Q_{L_3}} - \mu_{u_{R_3}} + \mu_H = 0 \label{eq:up_yukawa}\,,
\eal
where $\mu_{Q_{L_i}}$, $\mu_{u_{R_i}}$, $\mu_{d_{R_i}}$, 
$\mu_{\ell_{L_i}}$, $\mu_{\ell_{R_i}}$, $\mu_H$ are the
chemical potentials of $Q_{L_i}, u_{R_i}, d_{R_i}, \ell_{L_i}, \ell_{R_i},\,H$, respectively and $i = 1 \cdots 3$.

The hypercharge neutrality condition gives
\bal
&4(\mu_{u_{R_1}} + \mu_{u_{R_2}} + \mu_{u_{R_3}}) +
2(\mu_{Q_{L_1}} + \mu_{Q_{L_2}} + \mu_{Q_{L_3}}) - 
2(\mu_{\ell_{L_1}} + \mu_{\ell_{L_2}} + \mu_{\ell_{L_3}}) \nn \\
&-2(\mu_{d_{R_1}} + \mu_{d_{R_2}} + \mu_{d_{R_3}}) -
2(\mu_{\ell_{R_1}} +\mu_{\ell_{R_2}}+\mu_{\ell_{R_3}})+
4 \mu_H= 0\,\,.
\label{eq:hyper_charge_balance}
\eal
An electroweak sphaleron populates an equal amount of left chiral quarks in each generation. 
Hence, the baryon number is equal in each generation, and it implies
\bal
2 \mu_{Q_{L_1}} + \mu_{u_{R_1}} + \mu_{d_{R_1}}
=2 \mu_{Q_{L_2}} + \mu_{u_{R_2}} + \mu_{d_{R_2}}
=2 \mu_{Q_{L_3}} + \mu_{u_{R_3}} + \mu_{d_{R_3}}\,\,.
\label{eq:deltaB_i}
\eal
The equilibrium of the electroweak and QCD sphaleron gives
the following conditions.
\bal
&\mu_{\ell_{L_1}} + \mu_{\ell_{L_2}} + \mu_{\ell_{L_3}} + 
3 (\mu_{Q_{L_1}} + \mu_{Q_{L_2}} + \mu_{Q_{L_3}}) = 0\,\,, \label{eq:EW_sph}\\
&2(\mu_{Q_{L_1}} + \mu_{Q_{L_2}} + \mu_{Q_{L_3}}) - 
(\mu_{u_{R_1}} + \mu_{u_{R_2}} + \mu_{u_{R_3}}) - 
(\mu_{d_{R_1}} + \mu_{d_{R_2}} + \mu_{d_{R_3}}) = 0\,\,.
\label{eq:QCD_sph}
\eal
Note that Eq.\,\eqref{eq:QCD_sph} is not an independent constraint since it can be derived by summing Eqs.\,\eqref{eq:down_yukawa} and \eqref{eq:up_yukawa}.

Thus, we have 13 constraint equations and they are given from Eqs.\,\eqref{eq:lepton_yukawa}--\eqref{eq:EW_sph}. 
Solving these equations, we can write 
$\mu_{Q_{L_i}}$, $\mu_{u_{R_i}}$, $\mu_{d_{R_i}}$, $\mu_{\ell_{R_i}}$, $\mu_H$ in terms of $\mu_{\ell_{L_i}}$ as
\bal
&\mu_{Q_{L_1}} = \mu_{Q_{L_2}} = \mu_{Q_{L_3}} = -\dfrac{1}{9}(\mu_{\ell_{L_1}} + \mu_{\ell_{L_2}} + \mu_{\ell_{L_3}})\,,\nn\\
&\mu_{u_{R_1}} = \mu_{u_{R_2}} = \mu_{u_{R_3}} = \dfrac{5}{63}(\mu_{\ell_{L_1}} + \mu_{\ell_{L_2}} + \mu_{\ell_{L_3}})\,,\nn\\
&\mu_{d_{R_1}} = \mu_{d_{R_2}} = \mu_{d_{R_3}} = -\dfrac{19}{63}(\mu_{\ell_{L_1}} + \mu_{\ell_{L_2}} + \mu_{\ell_{L_3}})\,,\nn\\
&\mu_{\ell_{R_1}} = \dfrac{1}{21}(17 \mu_{\ell_{L_1}} - 4 \mu_{\ell_{L_2}} - 4 \mu_{\ell_{L_3}})\,,~~~~
\mu_{\ell_{R_2}} = \dfrac{1}{21}(-4 \mu_{\ell_{L_1}} + 17 \mu_{\ell_{L_2}} - 4 \mu_{\ell_{L_3}})\,,~~~~
\mu_{\ell_{R_3}} = \dfrac{1}{21}(-4 \mu_{\ell_{L_1}} - 4 \mu_{\ell_{L_2}} + 17 \mu_{\ell_{L_3}})\,\nn\\
&\mu_H = \dfrac{4}{21}(\mu_{\ell_{L_1}} + \mu_{\ell_{L_2}} + \mu_{\ell_{L_3}})\,\,,
\label{eq:mu_sol}
\eal
and the sphaleron factor is
\bal
a_{\rm sph} = \dfrac{n_{\Delta B}}{n_{\Delta B} - n_{\Delta L}} = \dfrac{28}{79}\,\,,
\eal
where
\bal
&n_{\Delta B} = \left(3 \times \dfrac{1}{3}\right) \dfrac{T^2}{6} 
\left[2(\mu_{Q_{L_1}} + \mu_{Q_{L_2}} + \mu_{Q_{L_3}})+
(\mu_{u_{L_1}} + \mu_{u_{L_2}} + \mu_{u_{L_3}}) + 
(\mu_{d_{R_1}} + \mu_{d_{R_2}} + \mu_{d_{R_3}})
\right]\,\,,\nn\\
&n_{\Delta L} = \dfrac{T^2}{6} 
\left[2(\mu_{\ell_{L_1}} + \mu_{\ell_{L_2}} + \mu_{\ell_{L_3}})+
(\mu_{\ell_{R_1}} + \mu_{\ell_{R_2}} + \mu_{\ell_{R_3}}) 
\right]\,\,.
\eal
Now, we define $n_{\Delta_\alpha} \equiv \dfrac{n_{\Delta B}}{3} - n_{\Delta L_\alpha}$ where 
$n_{\Delta L_\alpha} = T^2 (2 \mu_{\ell_{L_\alpha}} + \mu_{\ell_{R_\alpha}})/6$ and $\alpha = 1\cdots3$.
Using Eq.\,\eqref{eq:mu_sol}, we can write
\bal
n_{\Delta_\alpha} = \dfrac{T^2}{6} {\cal B}_{\alpha \beta} \mu_{\ell_{L_\beta}} \nn\\
= \dfrac{{\cal B}_{\alpha \beta}}{2} n_{\Delta \ell_{L_\beta}} \,\,,\nn\\
\eal
and in the last step, we use Eq.\,\eqref{eq:asymmetry_th_eq}. 
Dividing both sides by the entropy density $s$ gives
$Y_{\Delta_\alpha} = {\cal P}_{\alpha \beta} Y_{\Delta \ell_{L_\beta}}$
where ${P}_{\alpha \beta} = {\cal B}_{\alpha \beta}/2$. The explicit form of 
$\cal P$ is
\bal
\dfrac{1}{106}\begin{pmatrix}
    -205 & -8 & -8 \\
    -8 & -205 & -8\\
    -8 & -8 &  -205\,\,.
\end{pmatrix}
\label{eq:B_matrix}
\eal
The ${\cal A}$ matrix is defined in Eq.\,\eqref{eq:A_matrix} as -${\cal P}^{-1}$, and it is given by
\bal
\dfrac{1}{711}\begin{pmatrix}
    442 & -32 & -32 \\
    -32 & 442 & -32\\
    -32 & -32 & 442
\end{pmatrix}\,\,.
\label{eq:A_matrixform}
\eal
\section{RELEVANT CROSS SECTIONS FOR LEPTOGENISIS}
\label{appB:cross_sections}
In this section, the analytical form of the cross section is provided 
in terms of $x \equiv \hat{s}/M_1^2$ where $\hat{s}$ is the Mandelstam variable, $r_{\rm BL} = m_{\zbl}/M_1$, and
$r_{\eta} = m_{\eta}/M_1$.
\begin{enumerate}
\item \underline{$\ni \ni \to \bar{f} f$:}

\begin{align}
& \sigma_{\ni \ni \to \bar{f} f}
 = \dfrac{\gbl^4 {\qbl^{(f)}}^2 }{3 \pi M_1^2}
 \dfrac{\sqrt{x(x-4)}} {\left(x - \rbl^2 \right)^2
 + \rbl^2 (\Gamma_{\zbl}/M_1)^2}
 \,\,.
 \label{eq: process_1}
\end{align}
\item \underline{$\ni \ni \to \eta \eta:$}
\begin{align}
\sigma_{\ni \ni \to \eta \eta} &= 
36 \lambda_\phi^2 \vbl^2 \yf^2 T^{ss}_{\ni \ni \to \eta \eta} + 
\yf^4 \left(
T^{tt}_{\ni \ni \to \eta \eta} + T^{uu}_{\ni \ni \to \eta \eta} 
\right) + 
6 \lamf \vbl \yf^3\left(T^{ts}_{\ni \ni \to \eta \eta} + T^{us}_{\ni \ni \to \eta \eta}
\right) +\nn\\
&+ \yf^4 T^{tu}_{\ni \ni \to \eta \eta}\,\,,
\end{align}
where
\bal
T^{ss}_{\ni \ni \to \eta \eta} &= \dfrac{\sqrt{x-4} \sqrt{x-4 \rf^2}}{8 \pi  M_1^4 x \left(\rf^2-x\right)^2}\,\,,\\
T^{tt}_{\ni \ni \to \eta \eta} &= 
\frac{\left(2 \rf^2-x-8\right) \log \left(\frac{-\sqrt{x-4} \sqrt{x-4 \rf^2}-2 \rf^2+x}{\sqrt{x-4} \sqrt{x-4 \rf^2}-2 \rf^2+x}\right)-\frac{\sqrt{x-4} \sqrt{x-4 \rf^2} \left(2 \left(\rf^4-6 \rf^2+8\right)+x\right)}{\rf^4-4 \rf^2+x}}{8 \pi  M_1^2 x(x-4) }\,\,,\\
T^{uu}_{\ni \ni \to \eta \eta} &=
\frac{\left(2 \rf^2-x-8\right) \log \left(\frac{-\sqrt{x-4} \sqrt{x-4 \rf^2}-2 \rf^2+x}{\sqrt{x-4} \sqrt{x-4 \rf^2}-2 \rf^2+x}\right)-\frac{\sqrt{x-4} \sqrt{x-4 \rf^2} \left(2 \left(\rf^4-6 \rf^2+8\right)+x\right)}{\rf^4-4 \rf^2+x}}{8 \pi  M_1^2 x(x-4) }\,\,\\
T^{su}_{\ni \ni \to \eta \eta} &= \frac{\left(2 \rf^2+x-8\right) \log \left(\frac{-\sqrt{x-4} \sqrt{x-4 \rf^2}-2 \rf^2+x}{\sqrt{x-4} \sqrt{x-4 \rf^2}-2 \rf^2+x}\right)-2 \sqrt{x-4} \sqrt{x-4 \rf^2}}{4 \pi  M_1^3 x(x-4) x \left(x-\rf^2\right)}\,\,,\\
T^{st}_{\ni \ni \to \eta \eta}  & = \frac{\left(2 \rf^2+x-8\right) 
\log \left(\frac{-\sqrt{x-4} \sqrt{x-4 \rf^2} - 2 \rf^2+x}{\sqrt{x-4} \sqrt{x-4 \rf^2}-2 \rf^2+x}\right)
-2\sqrt{x-4} \sqrt{x-4 \rf^2}}{4 \pi M_1^3 x(x-4) \left(x-\rf^2\right)}\,\,,\\
T^{ut}_{\ni \ni \to \eta \eta}  & =
\frac{\left(\rf^4+4 (x-4)\right) \log \left(\left(\frac{\sqrt{x-4} \sqrt{x-4 \rf^2}-2 \rf^2+x}{\sqrt{x-4} \sqrt{x-4 \rf^2}+2 \rf^2-x}\right)^2\right)-\sqrt{x-4} 
\sqrt{x-4 \rf^2} \left(x-2 \rf^2\right)}{4 \pi  M_1^2 (x-4) x \left(x-2 \rf^2\right)}\,\,.
\eal
\item \underline{$\ni \ni \to \zbl \zbl:$}
\bal
\sigma_{\ni \ni \to \zbl \zbl} &= \gbl^4 \left(T^{tt}_{\ni \ni \to \zbl \zbl} + 
T^{uu}_{\ni \ni \to \zbl \zbl} + T^{ut}_{\ni \ni \to \zbl \zbl}\right) + \nn\\
&64 \gbl^4 \vbl^2 \yf^2 T^{ss}_{\ni \ni \to \zbl \zbl}
+8 \gbl^4 \vbl \yf \left(T^{st}_{\ni \ni \to \zbl \zbl} + T^{su}_{\ni \ni \to \zbl \zbl}\right)\,\,,
\eal
where
\allowdisplaybreaks
\bal
&T^{tt}_{\ni \ni \to \zbl \zbl} = \dfrac{1}{{48 \pi  M_1^2 x}}
\left[
\frac{\sqrt{x-4 \rbl^2} \left(-48 \rbl^8+4 \rbl^6 (5 x+46)+\rbl^4 ((x-102) x+32)\right)}{\rbl^4 \sqrt{x-4} \left(\rbl^4-4 \rbl^2+x\right)}
\right.\nn\\
&\left. +\frac{\sqrt{x-4 \rbl^2}\left(16 \rbl^2 (x-7) x+x^2 (x+2)\right)}{\rbl^4 \sqrt{x-4} \left(\rbl^4-4 \rbl^2+x\right)}
-\frac{24 \left(-2 \rbl^2+x+8\right) \log \left(\frac{-\sqrt{x-4} \sqrt{x-4 \rbl^2}-2 \rbl^2+x}{\sqrt{x-4} \sqrt{x-4 \rbl^2}-2 \rbl^2+x}\right)}{x-4}
\right]\,\,,\nn\\
&T^{uu}_{\ni \ni \to \zbl \zbl} = \dfrac{1}{{48 \pi  M_1^2 x}}  = \dfrac{1}{48 \pi M_1^2 x}
\left[
\frac{\sqrt{x-4 \rbl^2} \left(-48 \rbl^8+4 \rbl^6 (5 x+46)+\rbl^4 ((x-102) x+32)\right)}{\rbl^4 \sqrt{x-4} \left(\rbl^4-4 \rbl^2+x\right)}
\right.\nn\\
&\left.+
\frac{\sqrt{x-4 \rbl^2}\left(16 \rbl^2 (x-7) x+x^2 (x+2)\right)}{\rbl^4 \sqrt{x-4} \left(\rbl^4-4 \rbl^2+x\right)}-\frac{24 \left(-2 \rbl^2+x+8\right) \log \left(\frac{-\sqrt{x-4} \sqrt{x-4 \rbl^2}-2 \rbl^2+x}{\sqrt{x-4} \sqrt{x-4 \rbl^2}-2 \rbl^2+x}\right)}{x-4}
\right]\,\,,\nn\\
&T^{ut}_{\ni \ni \to \zbl \zbl} = 
-\frac{\frac{\left(4 \rbl^2 (5 x+22)+(x-46) x\right) \sqrt{x-4 \rbl^2}}{\sqrt{x-4} x}+\frac{48 \left(\rbl^2-1\right) \left(\rbl^4-4 \rbl^2+x\right) \log \left(\frac{\left(\sqrt{x-4} \sqrt{x-4 \rbl^2}-2 \rbl^2+x\right)^2}{\left(\sqrt{x-4} \sqrt{x-4 \rbl^2}+2 \rbl^2-x\right)^2}\right)}{(x-4) \left(2 \rbl^2-x\right)}}{24 \pi  M_1^2 \rbl^4}\,\,,\nn\\
&T^{ss}_{\ni \ni \to \zbl \zbl} = 
\frac{\sqrt{x-4} \sqrt{x-4 \rbl^2} \left(12 \rbl^4-4 \rbl^2 x+x^2\right)}{32 \pi  M_1^4 \rbl^4 x \left(x-\rf^2\right)^2}\,\,,\nn\\
&T^{st}_{\ni \ni \to \zbl \zbl} = 
\frac{1}{4 \pi  \text{M1}^3 \rbl^4 (x-4) x \left(x-\rf^2\right)}
\left[
-\sqrt{x-4} \sqrt{x-4 \rbl^2} \left(4 \rbl^4-2 \rbl^2 x+x^2\right) \right.\nn\\
&\left.+2 \left(2 \rbl^6-8 \rbl^4+4 \rbl^2 x-x^2\right) \log \left(\frac{-\sqrt{x-4} \sqrt{x-4 \rbl^2}-2 \rbl^2+x}{\sqrt{x-4} \sqrt{x-4 \rbl^2}-2 \rbl^2+x}\right)
\right]\,\,,\nn\\
&T^{su}_{\ni \ni \to \zbl \zbl} = 
\frac{1}{4 \pi  M_1^3 \rbl^4 (x-4) x \left(x-\rf^2\right)}
\left[
-\sqrt{x-4} \sqrt{x-4 \rbl^2} \left(4 \rbl^4-2 \rbl^2 x+x^2\right) + \right.\nn\\
&\left.2 \left(2 \rbl^6-8 \rbl^4+4 \rbl^2 x-x^2\right) \log \left(\frac{-\sqrt{x-4} \sqrt{x-4 \rbl^2}-2 \rbl^2+x}{\sqrt{x-4} \sqrt{x-4 \rbl^2}-2 \rbl^2+x}\right)\right]\,\,.
\eal
\item \underline{$\ni \ni \to \eta \zbl:$}
\bal
\sigma_{\ni \ni \to \eta \zbl}&= \gbl^2 \yf^2 \left(T^{tt}_{\ni \ni \to \eta \zbl} + T^{tt}_{\ni \ni \to \eta \zbl}
T^{ut}_{\ni \ni \to \eta \zbl}\right) + 
64 \gbl^6 \vbl^2 T^{ss}_{\ni \ni \to \eta \zbl} + \nn\\
&8 \gbl^4 \yf \vbl (T^{st}_{\ni \ni \to \eta \zbl} +
 T^{su}_{\ni \ni \to \eta \zbl})\,\,,
\eal
where
\bal
&T^{tt}_{\ni \ni \to \eta \zbl} = \dfrac{1}{16\pi M_1^2 x (x-4)}
\left[
4 \left(\rbl^2+\rf^2-x-8\right) \log \left(-\frac{\sqrt{x-4} \sqrt{\frac{\left(\rbl^2-\rf^2\right)^2}{x}-2 \left(\rbl^2+\rf^2\right)+x}+\rbl^2+\rf^2-x}{\sqrt{x-4} \sqrt{\frac{\left(\rbl^2-\rf^2\right)^2}{x}-2 \left(\rbl^2+\rf^2\right)+x}-\rbl^2-\rf^2+x}\right)\right.\nn\\
&\left.
+\frac{2 \sqrt{x-4} \sqrt{\frac{\rbl^4-2 \rbl^2 \left(\rf^2+x\right)+\left(\rf^2-x\right)^2}{x}} \left(-2 \rbl^6+\rbl^4 \left(-4 \rf^2 (x-1)+13 x-4\right)-\rbl^2 \left(\rf^2-4\right) \left(2 \rf^2-x (x+6)\right)\right)}{\rbl^2 \left(\rbl^4+\rbl^2 \left(\rf^2 (x-2)-2 x\right)+\left(\rf^2-x\right)^2\right)} \right.\nn\\
&\left.
+\frac{2 \sqrt{x-4} \sqrt{\frac{\rbl^4-2 \rbl^2 \left(\rf^2+x\right)+\left(\rf^2-x\right)^2}{x}}\left(
(x-4) \left(\rf^2-x\right)^2\right)}{\rbl^2 \left(\rbl^4+\rbl^2 \left(\rf^2 (x-2)-2 x\right)+\left(\rf^2-x\right)^2\right)}
\right]\,\,,\\
& T^{uu}_{\ni \ni \to \eta \zbl} = \dfrac{1}{16 \pi M_1^2 x(x-4)}
\left[
4 \left(\rbl^2+\rf^2-x-8\right) \log \left(-\frac{\sqrt{x-4} \sqrt{\frac{\left(\rbl^2-\rf^2\right)^2}{x}-2 \left(\rbl^2+\rf^2\right)+x}+\rbl^2+\rf^2-x}{\sqrt{x-4} \sqrt{\frac{\left(\rbl^2-\rf^2\right)^2}{x}-2 \left(\rbl^2+\rf^2\right)+x}-\rbl^2-\rf^2+x}\right)  \right.\nn\\
&\left.
+\frac{2 \sqrt{x-4} \sqrt{\frac{\rbl^4-2 \rbl^2 \left(\rf^2+x\right)+\left(\rf^2-x\right)^2}{x}} \left(-2 \rbl^6+\rbl^4 \left(-4 \rf^2 (x-1)+13 x-4\right)-\rbl^2 \left(\rf^2-4\right) \left(2 \rf^2-x (x+6)\right)\right)}{\rbl^2 \left(\rbl^4+\rbl^2 \left(\rf^2 (x-2)-2 x\right)+\left(\rf^2-x\right)^2\right)} \right.\nn\\
&\left.
+\frac{
2 \sqrt{x-4} \sqrt{\frac{\rbl^4-2 \rbl^2 \left(\rf^2+x\right)+\left(\rf^2-x\right)^2}{x}}
\left((x-4) \left(\rf^2-x\right)^2\right)}{\rbl^2 \left(\rbl^4+\rbl^2 \left(\rf^2 (x-2)-2 x\right)+\left(\rf^2-x\right)^2\right)}
\right]\,\,,\\\nn
& T^{ut}_{\ni \ni \to \eta \zbl} = \dfrac{1}{4 \pi M_1^2 \rbl^2 (x-4) x \left(\rbl^2+\rf^2-x\right)}\left[
- \sqrt{x-4} \left(-2 \rbl^2+x+4\right) \left(-\rbl^2-\rf^2+x\right) \times \right.\nn\\
&\left.\sqrt{\frac{\rbl^4-2 \rbl^2 \left(\rf^2+x\right)+\left(\rf^2-x\right)^2}{x}} 
+2  \left(\rbl^4 \left(\rf^2-1\right)+\rbl^2 \left(-\left(\rf^2 (x-2)\right)+6 x-16\right)-\left(\rf^2-x\right)^2\right) \times 
\right.\nn\\
&\left.
\log \left(\frac{\left(\sqrt{x-4} \sqrt{\frac{\left(\rbl^2-\rf^2\right)^2}{x}-2 \left(\rbl^2+\rf^2\right)+x}+\rbl^2+\rf^2-x\right)^2}{\left(-\sqrt{x-4} \sqrt{\frac{\left(\rbl^2-\rf^2\right)^2}{x}-2 \left(\rbl^2+\rf^2\right)+x}+\rbl^2+\rf^2-x\right)^2}\right)
\right]\,\,\nn\\
& T^{ss}_{\ni \ni \to \eta \zbl} = \frac{\sqrt{\frac{\rbl^4-2 \rbl^2 \left(\rf^2+x\right)+\left(\rf^2-x\right)^2}{x}}}{48 \pi x^2 M_1^4 \rbl^6 (x-\rbl^2)^2 \sqrt{x-4}}
\left[
\rbl^8 (x+2)-12 \rbl^2 x \left(\rf^4-\rf^2 x+2 x^2\right)+6 x^2 \left(\rf^2-x\right)^2\right.\nn\\
&\left.-2 \rbl^6 \left(\rf^2 (x+2)+x (32-5 x)\right)+\rbl^4 \left(\rf^4 (x+2)-2 \rf^2 (x-10) x+x^2 (x+32)\right)
\right]\,\,,\nn\\
& T^{st}_{\ni \ni \to \eta \zbl} = \dfrac{1}{4 \pi  M_1^3 \rbl^4 (x-4) x \left(\rbl^2-x\right)}
\left[
\sqrt{x-4} \left(4 \rbl^4-x \left(\rbl^2+\rf^2\right)+x^2\right) \times \right.\nn\\
&\left.
\sqrt{\frac{\rbl^4-2 \rbl^2 \left(\rf^2+x\right)+\left(\rf^2-x\right)^2}{x}} +
2 \left(\rbl^6+\rbl^4 \left(\rf^2+x-9\right)+\rbl^2 \left(2 x-\rf^2 (x-2)\right)-\left(\rf^2-x\right)^2\right) \times\right.\nn\\
&\left.
\log \left(\frac{-\sqrt{x-4} \sqrt{\frac{\left(\rbl^2-\rf^2\right)^2}{x}-2 \left(\rbl^2+\rf^2\right)+x}+\rbl^2+\rf^2-x}{\sqrt{x-4} \sqrt{\frac{\left(\rbl^2-\rf^2\right)^2}{x}-2 \left(\rbl^2+\rf^2\right)+x}+\rbl^2+\rf^2-x}\right)
\right]\,\,,\nn\\
& T^{su}_{\ni \ni \to \eta \zbl} = 
\dfrac{1}{4 \pi  M_1^3 \rbl^4 (x-4) x \left(\rbl^2-x\right)}
\left[
\sqrt{x-4} \left(4 \rbl^4-x \left(\rbl^2+\rf^2\right)+x^2\right) \times \right.\nn\\
&\left. \sqrt{\frac{\rbl^4-2 \rbl^2 \left(\rf^2+x\right)+\left(\rf^2-x\right)^2}{x}}
-2 \left(\rbl^6+\rbl^4 \left(\rf^2+x-9\right)+\rbl^2 \left(2 x-\rf^2 (x-2)\right)-\left(\rf^2-x\right)^2\right) \times \right.\nn\\
&\left.
\log \left(-\frac{\sqrt{x-4} \sqrt{\frac{\left(\rbl^2-\rf^2\right)^2}{x}-2 \left(\rbl^2+\rf^2\right)+x}+\rbl^2+\rf^2-x}{\sqrt{x-4} \sqrt{\frac{\left(\rbl^2-\rf^2\right)^2}{x}-2 \left(\rbl^2+\rf^2\right)+x}-\rbl^2-\rf^2+x}\right)
\right]\,\,.
\eal
\item \underline{$\bar{t}_R U_{3_L} \to \ell_{\alpha_L} \ni$:}
\bal
\sigma_{\bar{t}_R U_{3_L} \to \ell_{\alpha_L} \ni} = 
|Y_{33}|^2 |{\Yd}_{\alpha i} |^2\frac{(x-1)^2}{4 \pi  M_1^2 x^3}\,\,.
\eal
\item \underline{$\ell_{\alpha_L} t_R \to U_{3_L} \ni$:}
\bal
\sigma_{\ell_{\alpha_L} t_R \to U_{3_L} \ni} = |Y_{33}|^2 |{\Yd}_{\alpha i} |^2
\frac{(x-1) \left(2 \delta ^2+x-2\right)+\left(2 \delta ^2-1\right) \left(\delta ^2+x-1\right) \log \left(\frac{\delta ^2}{\delta ^2+x-1}\right)}{4 \pi  M_1^2 x^2 \left(\delta ^2+x-1\right)}\,\,.
\eal
\item \underline{$\bar{\ell}_{\alpha_L} U_{3_L} \to t_R \ni$:}
\bal
\sigma_{\bar{\ell}_{\alpha_L} U_{3_L} \to t_R \ni}=|Y_{33}|^2 |{\Yd}_{\alpha i} |^2
\frac{(x-1) \left(2 \delta ^2+x-2\right)+\left(2 \delta ^2-1\right) \left(\delta ^2+x-1\right) \log \left(\frac{\delta ^2}{\delta ^2+x-1}\right)}{4 \pi  M_1^2 x^2 \left(\delta ^2+x-1\right)}\,\,.
\eal
In both of the above equations, $Y_{33}$ is the third-generation up-type Yukawa coupling, and we added a 
regulator mass $\delta M_1$ in the $t$ channel propagator, and we consider
$\delta = 0.1$ throughout our analysis.
\end{enumerate}
\section{COLLISION TERMS FOR LEPTON-FLAVOR-VIOLATING PROCESSES}
\label{appC:LFV}
There are four scatterings involving the members of the left-handed
lepton doublet and Higgs doublet where the lepton flavor ($\ell_\alpha$)
is violated. These scatterings and corresponding collision terms are listed below.
\begin{itemize}
\item $\ell_{\alpha} H \rightleftharpoons \overline{\ell_{\beta}} H^{\star}$ \\
\begin{eqnarray}
\mathcal{C}_{\alpha\beta} =
-2 s \sum_{\beta} \langle {\sigma^{\prime}_{\ell_{\alpha} H \rightarrow \overline{\ell_{\beta}} H^{\star}} v} \rangle\,Y^{\rm eq}_{\ell_\alpha}
\,Y^{\rm eq}_{H}\,\,\left(\dfrac{n_{\Delta{\ell}_{\alpha}}}{2\,n^{\rm eq}_{\alpha}}
+ \dfrac{n_{\Delta{\ell}_{\beta}}}{2\,n^{\rm eq}_{\beta}} \right)\,,
\end{eqnarray}
\item $\ell_{\alpha} \ell_{\beta} \rightleftharpoons
H^{\star} H^{\star}$ \\
\begin{eqnarray}
\mathcal{C}_{\alpha\beta} =
-2s\sum_{\beta} \langle {\sigma_{\ell_{\alpha} \ell_{\beta} \rightarrow H^{\star} H^{\star}} v} \rangle\,Y^{\rm eq}_{\ell_\alpha}
\,Y^{\rm eq}_{H}\,\,\left(
\dfrac{n_{\Delta{\ell}_{\alpha}}}{2\,n^{\rm eq}_{\alpha}}
+
\dfrac{n_{\Delta{\ell}_{\beta}}}{2\,n^{\rm eq}_{\beta}}
\right)\,,
\end{eqnarray}

\item $\ell_{\alpha} \overline{\ell_{\beta}} \rightleftharpoons
H H^{\star}$ ($\alpha \neq \beta$) \\
\begin{eqnarray}
\mathcal{C}_{\alpha\beta} =
-2s\sum_{\beta} \langle {\sigma_{\ell_{\alpha} \overline{\ell_{\beta}} \rightarrow H H^{\star}} v} \rangle\,Y^{\rm eq}_{\ell_\alpha}
\,Y^{\rm eq}_{H}\,\,\left(
\dfrac{n_{\Delta{\ell}_{\alpha}}}{2\,n^{\rm eq}_{\alpha}}
-
\dfrac{n_{\Delta{\ell}_{\beta}}}{2\,n^{\rm eq}_{\beta}}
\right)\,,
\end{eqnarray}
\item $\ell_{\alpha} H \rightleftharpoons
\ell_{\beta} H $ ($\alpha \neq \beta$) \\
\begin{eqnarray}
\mathcal{C}_{\alpha\beta} =
-2s\sum_{\beta} \langle {\sigma_{\ell_{\alpha} H \rightarrow
\ell_{\beta} H } v} \rangle\,Y^{\rm eq}_{\ell_\alpha}
\,Y^{\rm eq}_{H}\,\,\left(
\dfrac{n_{\Delta{\ell}_{\alpha}}}{2\,n^{\rm eq}_{\alpha}}
-
\dfrac{n_{\Delta{\ell}_{\beta}}}{2\,n^{\rm eq}_{\beta}}
\right)\,,
\end{eqnarray}
\end{itemize}
\section{FIELD-DEPENDENT MASSES AND THERMAL MASSES}
\label{AppD:masses}
The field-dependent mass of the scalar fields in the diagonal basis i.e., in the $(\xi, \,\eta)$ basis, is given by 
\bal
&\mbar_{\zeta}^2 = \dfrac{\mbar_{\zeta_1}^2 + \mbar_{\eta_1}^2 + \sqrt{(\mbar_{\zeta_1}^2-\mbar_{\eta_1}^2)^2 + 4 \mbar_{\zeta_1 \eta_1}^4}}{2}\,\,\nn\\
&\mbar_{\eta}^2 = \dfrac{\mbar_{\zeta_1}^2 + \mbar_{\eta_1}^2 - \sqrt{(\mbar_{\zeta_1}^2-\mbar_{\eta_1}^2)^2 + 4 \mbar_{\zeta_1 \eta_1}^4}}{2}\,\,,
\eal
where 
    \bal
        \label{eq:field_dep_mass_scalar}
        &\mbar_{\zeta_1}^2      = -\mu_H^2 + 3 \lambda_H \zeta_1^2 - \dfrac{\lambda^\prime}{2}\eta_1^2 \,,~~~~
        \mbar_{\eta_1}^2   = -\mu_\Phi^2 + 3 \lambda_\Phi \eta_1^2 - \dfrac{\lambda^\prime}{2}\zeta_1^2\,,~~~~
        \mbar^2_{\zeta_1 \eta_1} = -\lambda^\prime \zeta_1 \eta_1
        \eal
    
    The field-dependent mass of the neutral SM gauge boson is given by 
    \bal
    \label{eq:field_dep_mass_gauge_boson}
    &\mbar_Z^2 = \dfrac{\mbar^2_{11} + \mbar^2_{22} + \sqrt{(\mbar^2_{11} - \mbar^2_{22})^2 + 4 \mbar^4_{12}}}{2}\,\,,\nn\\
    & \mbar_{\gamma}^2 = \dfrac{\mbar^2_{11} + \mbar^2_{22} - \sqrt{(\mbar^2_{11} - \mbar^2_{22})^2 + 4 \mbar^4_{12}}}{2}\,\,,
    \eal
    where
    \bal
    &\mbar_{11}^2 = \dfrac{g_2^2 \zeta_1^2}{4},~~~~~~ \mbar^2_{22} = \dfrac{g_1^2 \zeta_1^2}{4},~~~~~\mbar^2_{12}=\dfrac{-g_1 g_2 \zeta_1^2}{4}\,\,.
    \eal

    The field-dependent mass of the Goldstone modes, $W$ boson, $\zbl$, and fermions are as follows:
        \bal
        &\mbar_{G^+}^2   = -\mu_H^2 +  \lambda_H \zeta_1^2 - \dfrac{\lambda^\prime}{2}\eta_1^2\,,~~~~
        \mbar_{\zeta_2}^2   = -\mu_H^2 +  \lambda_H \zeta_1^2 - \dfrac{\lambda^\prime}{2}\eta_1^2\,,~~~~
        \mbar_{\eta_2}^2   = -\mu_\Phi^2 +  \lambda_\Phi \eta_1^2 - \dfrac{\lambda^\prime}{2}\zeta_1^2\,,\nn\\
        &\mbar_W^2      = \left(\dfrac{m_W}{\vh}\right)^2 \zeta_1^2 \,,~~~~
        \mbar_{\zbl}^2 = \left(\dfrac{m_{\zbl}}{\vbl}\right)^2 \eta_1^2\,,~~~~
        \mbar_t^2      = \left(\dfrac{m_t}{\vh}\right)^2 \zeta_1^2 \,,~~~~
        \mbar_{i}^2  = \left(\dfrac{M_{i}}{\vbl}\right)^2 \eta_1^2\,.
    \eal
One can obtain the field-dependent mass of $\xi$ and $\eta$ in the thermal background by performing the following replacement in Eq.\,\eqref{eq:field_dep_mass_scalar}:
\bal
\mbar^2_{\zeta} \to \mbar_{\zeta}^2 + \Pi^2_{\zeta_1},~~~~~~~~~ \mbar^2_{\eta} \to \mbar_{\eta}^2 + \Pi^2_{\eta_1}\,\,,
\eal
where
\bal
\label{eq:scalar_thermal_mass}
&\Pi^2_{\zeta_1} = \left(\dfrac{g_1^2}{16} + \dfrac{3 g_2^2}{16} + \dfrac{y_t^2}{4} + \dfrac{\lambda_H}{2} - \dfrac{\lambda^\prime}{12}\right)T^2\,\,,\nn\\
&\Pi^2_{\eta_1} = \left(\gbl^2 + \dfrac{3 g_2^2}{16} + \dfrac{\yf}{8} + \dfrac{\lambda_\Phi}{3} - \dfrac{\lambda^\prime}{6}\right)T^2\,\,. \nn\\
\eal
Similarly, the longitudinal modes of the gauge boson get thermal correction and $\mbar^{L}_{W,T} = \mbar_W + 11 g_2^2 T^2/6$. Similar to the scalar case, one can 
obtain the thermal mass correction to the longitudinal modes of the neutral SM gauge bosons by replacing $\mbar_{11}^2 \to \mbar^2_{11} + 11 g_2^2 T^2/6$ and 
$\mbar_{22}^2 \to \mbar^2_{22} + 11 g_1^2 T^2/6$ in Eq.\,\eqref{eq:field_dep_mass_gauge_boson}. The field-dependent mass of Goldstone modes and longitudinal $\zbl$
in the thermal background is as follows:
\bal
&\mbar_{G^+,T}^2 = \mbar^2_{G^+} + \Pi^2_{\zeta_1},~~~~~\mbar_{\zeta_2,T}^2 = \mbar^2_{\zeta_2} + \Pi^2_{\zeta_1},~~~~~~~ \mbar_{\eta_2,T}^2 = \mbar^2_{\eta_2} + \Pi^2_{\eta_1}\,,\nn\\
&\mbar^2_{\zbl,T} = \mbar_{\zbl}^2 + 4 \gbl^2 T^2\,\,.
\eal
\end{widetext}

\bibliography{ref}
\end{document}